\pdfoutput=1

\documentclass[11pt]{article}
\RequirePackage[OT1]{fontenc}
\RequirePackage{amsthm,amsmath}
\RequirePackage[numbers]{natbib}
\RequirePackage[colorlinks,citecolor=blue,urlcolor=blue]{hyperref}

\usepackage{indentfirst}
\usepackage[utf8]{inputenc} 
\usepackage[T1]{fontenc}    
\usepackage[colorlinks]{hyperref} 
\usepackage{url}            
\usepackage{booktabs}       
\usepackage{amsfonts}       
\usepackage{amsmath}
\usepackage{nicefrac}       
\usepackage{microtype}      
\usepackage{color}
\usepackage{bbm}
\usepackage{amssymb, enumerate}
\usepackage{makecell}
\usepackage{wrapfig}
\usepackage{extarrows}
\usepackage{arydshln}

\usepackage{caption}
\usepackage{multirow}

\usepackage{amsmath,amsfonts,amsthm, amssymb, enumerate}
\usepackage{lipsum}
\usepackage{caption}
\usepackage{amssymb,graphicx,subfigure}
\usepackage[bottom]{footmisc}
\usepackage[dvipsnames]{xcolor}
\usepackage{array,multirow}
\usepackage{enumitem}

\usepackage{microtype}
\usepackage{graphicx}
\usepackage{subfigure}
\usepackage{booktabs} 
 
\usepackage{graphicx}
\usepackage{amsmath}
\usepackage{amssymb}
\usepackage{booktabs}  
\usepackage{times}
\usepackage{epsfig}
\usepackage{graphicx}
\usepackage{amsfonts, amsthm, enumerate} 
\usepackage[bottom]{footmisc} 
\usepackage{enumitem}
\usepackage{setspace}
\usepackage{wrapfig}
\usepackage{extarrows}
\usepackage{makecell}
\usepackage{bbm}  
\usepackage{textcomp}
\usepackage{xcolor} 
\usepackage{caption}

\usepackage[font=small,labelfont=bf]{caption}

\usepackage{algorithm,algpseudocode}
\algnewcommand\TR{\item[{\textbf{Training phase}}]}
\algnewcommand\TE{\item[{\textbf{Test phase}}]}
\algnewcommand\Input{\item[{{Input:}}]}
\algnewcommand\Output{\item[{{Output:}}]}
\algnewcommand\Initialize{\item[{{Initialize:}}]}
\algnewcommand{\return}[1]{
	\State \textbf{return:}
	\Statex \hspace*{\algorithmicindent}\parbox[t]{.8\linewidth}{\raggedright #1}
}


\usepackage{setspace}
\begin{document}
	\title{Automated Precision Localization of Peripherally Inserted Central Catheter Tip through Model-Agnostic Multi-Stage Networks}
	\date{}
	\author{
		Subin Park$^{1}$\thanks{Equal contribution}, Yoon Ki Cha$^{2}$\footnotemark[1], Soyoung Park$^{1}$, Kyung-Su Kim$^{3,4}$\thanks{Corresponding author: Kyung-Su Kim (kskim.doc@gmail.com) and Myung Jin Chung (mj1.chung@samsung.com)}, Myung Jin Chung$^{2,3,4}$\footnotemark[2]\\ 
		\\
		{\footnotesize $^{1}$Department of Health Sciences and Technology, SAIHST, Sungkyunkwan University, Seoul 06351, Republic of Korea}\\
		{\footnotesize $^{2}$Department of Radiology, Samsung Medical Center, Sungkyunkwan University School of Medicine, Seoul, Republic of Korea}\\
		{\footnotesize $^{3}$Medical AI Research Center, Research Institute for Future Medicine, Samsung Medical Center, Seoul, Republic of Korea}\\
		{\footnotesize $^{4}$Department of Data Convergence and Future Medicine, Sungkyunkwan University School of Medicine, Seoul, Republic of Korea}
	} 
 
	\maketitle

\begin{abstract}
\noindent
\textbf{Background} $\,\,$
Peripherally inserted central catheters (PICCs) have been widely used as one of the representative central venous lines (CVCs) due to their long-term intravascular access with low infectivity. However, PICCs have a fatal drawback of a high frequency of tip mispositions, increasing the risk of puncture, embolism, and complications such as cardiac arrhythmias. To automatically and precisely detect it, various attempts have been made by using the latest deep learning (DL) technologies. However, even with these approaches, it is still practically difficult to determine the tip location because the multiple fragments phenomenon (MFP) occurs in the process of predicting and extracting the PICC line required before predicting the tip. \\
\textbf{Objective} $\,\,$
This study aimed to develop a system generally applied to existing models and to restore the PICC line more exactly by removing the MFs of the model output, thereby precisely localizing the actual tip position for detecting its misposition. \\
\textbf{Methods} $\,\,$
To achieve this, we proposed a multi-stage DL-based framework post-processing the PICC line extraction result of the existing technology. Our method consists of the following three stages: 1. Existing PICC line segmentation network for a baseline, 2. Patch-based PICC line refinement network, 3. PICC line reconnection network. The proposed second and third stage models address MFs caused by the sparseness of the PICC line and the line disconnection due to confusion with anatomical structures respectively, thereby enhancing the tip detection. \\
\textbf{Results} $\,\,$
To verify the objective performance of the proposed MFCN, internal validation and external validation were conducted. For internal validation, learning (130 samples) and verification (150 samples) were performed with 280 data including PICC among Chest X-ray (CXR) images taken at our institution. External validation was conducted using a public dataset called the Royal Australian and New Zealand college of radiologists (RANZCR), and training (130 samples) and validation (150 samples) were performed with 280 data of CXR images including PICC, which has the same number as that for internal validation. The performance was compared by each root mean squared error (RMSE) and MFP incidence rate (i.e., rate at which model predicts PICC as multiple sub-lines) according to whether or not MFCN is applied to five conventional models (i.e., FCDN, UNET, AUNET, FCDN-HT, and UNET-RPN). In internal validation, when MFCN was applied to the existing single model, MFP was improved by an average of 45 \%. The RMSE was improved by over 63\% from an average of 26.85mm (17.16 to 35.80mm) to 9.72mm (9.37 to 10.98mm). In external validation, when MFCN was applied, the MFP incidence rate decreased by an average of 32\% and the RMSE decreased by an average of 65\%. Therefore, by applying the proposed MFCN, we observed the consistent detection performance improvement of PICC tip location compared to the existing model.\\
\textbf{Conclusions} $\,\,$
In this study, we applied the proposed technique to the existing technique and demonstrated that it provides high tip detection performance, proving its high versatility and superiority. Therefore, we believe, in countries and regions where radiologists are scarce, the proposed DL approach will able to effective detect PICC misposition on behalf of radiologists.
\end{abstract}

\section{Introduction}
\label{Sec:Introduction}

Central venous catheter (CVC) lines are essential in the care of patients with severe diseases treated in surgical, intensive care and oncological units with fluid and transfusion therapy, including total parenteral nutrition, antibiotic therapy and also for chemotherapy purposes \cite{johansson2013advantages}.Traditionally, clinicians used non-tunneled or tunneled CVCs by centrally inserting catheters into the superior vena cava (SVC) via the subclavian or the internal jugular veins depending on the indication and on how long the patient will require the CVCs. Since direct puncture of one of the great veins at the upper thoracic aperture by CVCs are difficult to perform and carries serious complications such as pneumothorax and hemothorax \cite{maki2006risk}, peripherally inserted central catheter (PICC) was introduced to clinical practice for easy access, longer usage and fewer complications alternative to traditional CVCs \cite{hammarskjold2008peripherally}. The PICC is a thin and long flexible catheter made of biocompatible material, either silicone or polyurethane, inserted percutaneously into the basilic or cephalic vein in the forearm, often with the help of ultrasound or fluoroscopy guidance. The catheter is then advanced into the central circulation with tip of the catheter most often placed in the SVC or at the junction of the superior vena cava and the right atrium \cite{maki2006risk}.

Mispositioned or migration of PICCs can have potentially serious complications such as thrombus formation or cardiac arrhythmia \cite{funaki2002central}. As a result, PICC positioning is always confirmed with a chest radiograph (CXR) immediately after the insertion. This requires timely and accurate interpretation by a radiologist. Delays in treatment initiation can be substantial particularly when this radiograph is one of many in a long list waiting to be interpreted \cite{tomaszewski2017time}. Deep learning techniques, however, may help prioritize and triage the review of radiographs to the top of a radiologist’s queue, improving workflow and turnaround time.

If the misplaced tip of PICC can be automatically detected, the diagnosis time can be shortened and the complication rate can be reduced by repositioning or removing it. To achieve this, the existing studies tried to automatically track the whole line of PICC and its tip together with the aid of image processing, pattern recognition, and recently deep learning (DL)-based technology \cite{keller2007semi, yu2020detection, yi2020automatic, ambrosini2017fully, lee2018deep, subramanian2019automated}. The expert clinicians generally detect the position of the PICC tip by following the line from the relatively visible position of the PICC line on the outside to the inside of the lung. Following this doctor's diagnosis process, these aforementioned studies further strengthened the interpretability by allowing the model to predict the entire PICC line rather than simply predicting the tip position. In other words, these studies detected the tip based on the segmented shape after segmenting the whole PICC line. In particular, even with the latest DL-based technology, the PICC tip is very small so difficult to learn and provide exact location of the tip. Therefore, it is necessary to segment the entire PICC line in order to compensate for this inaccuracy issue of prediction result for the tip position.

\begin{figure}[t!]
    \centering
    \includegraphics[width=0.8\textwidth]{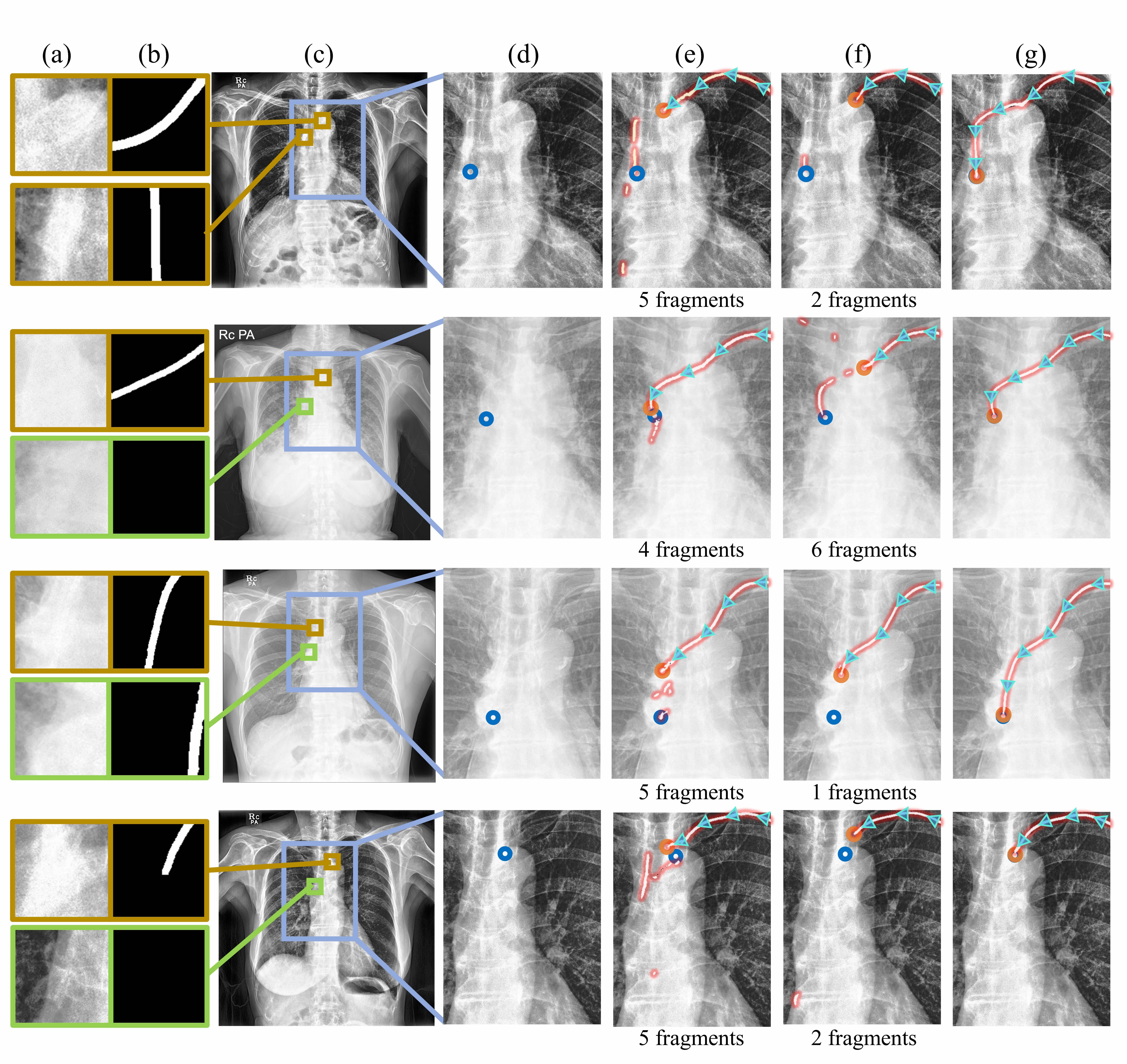}
    \caption{Illustration of MFPs and the corresponding tip detection problem of the existing AI models: (a) ROI of CXR image (b) Ground truth PICC line (c) Original CXR image (d) Ground truth tip (e) line prediction by FCN (f) line prediction by UNET (g) line prediction by proposed MFCN (blue arrow, used as tracking path of PICC tip). In the CXR image, if there exists an object whose shape is similar to that of PICC such as a spine and clavicle (green border image), or if the actual PICC line is shown as blurred due to being overlapped by some anatomical structures (brown border image), a false-negative (i.e., break) or false-positive error occurs in the PICC segmentation result. As the AI model tracks this PICC segmentation result (blue arrow) and selects the unbroken lowest point of it as the tip location (orange circle), so these errors provide a significant error in the precise location prediction of the actual tip.}
    \label{fig1:MFP}
\end{figure}

However, there still exist some technical limitations in accurately predicting the tip location even with the help of PICC full line segmentation results. The typical limitation is the multiple fragments phenomenon (MFP): In the process of predicting the PICC line by the model, it is not predicted as one complete line as the ground truth, but multiple divided lines are provided. These MFPs occur frequently even with the gold standard DL-based models such as FCN \cite{long2015fully} and U-Net (UNET) \cite{ronneberger2015u} as shown in Figure \ref{fig1:MFP}. This MFP makes the prediction model cause both false detection and non-detection of the actual PICC line, thereby making it difficult for the clinician to determine the actual tip position (i.e., distinguish where the correct end of the PICC catheter is) even with the aid of these automatic diagnosis models by tracking their PICC line estimates. 

Existing DL-based studies introduced the process of extracting the entire PICC line to effectively detect the tip, but in this process, the MFP was not directly solved. This makes it difficult to accurately identify the location of the tip in the extracted PICC line estimate, limiting its clinical use. Specifically, there have been studies to detect the tip position by dividing the entire structure of the PICC line \cite{ambrosini2017fully, subramanian2019automated}, but they still have the problem of MFP as they assume that the complete PICC line is extracted. Yu \textit{et al}. detected the tip position more precisely by performing object detection on the tip position separately in addition to PICC line segmentation \cite{yu2020detection}. And Yi \textit{et al}. tried to solve the difficulties of data annotation by creating a virtual PICC using the generated model \cite{yi2020automatic}. However, neither of these studies focused on directly solving the MFP to improve the performance of tip detection. The most relevant study to supplement MFP is that published by Lee \textit{et al}. \cite{lee2018deep}. They proposed a post-processing technique using the hough transform \cite{duda1972use, kiryati1991probabilistic} applied to the PICC line segmentation result and tried to resolve the MFP to improve the tip detection performance. However, these results are sensitive to the hough transform parameter, so it is difficult to have a common parameter in which MFP does not occur in all images.

Our study focused on the MFP problem of existing DL-based PICC tip detection models and proposed a method generally applied to these existing models to solve the MFP problem so consistently improve their performance of PICC tip detection. We call this method the multi-fragment complementary network (MFCN). The proposed MFCN consisted of the following three stages: Stage 1. Backbone network for PICC line segmentation, Stage 2. Patch-based PICC line segmentation network, Stage 3. PICC line reconnection network.

Stage 1 is an arbitrary network for PICC line segmentation, and any existing proposed model can be applied. That is, the proposed MFCN is the result solving the MFP problem of PICC line segmentation of Stage 1 by applying two models of Stage 2 and 3 at the next steps. Therefore, the main contribution of our study is in these Stage 2 and Stage 3 models.

Stage 2 model of MFCN, the patch-based segmentation model, extracts a small patch in a whole CXR image by using a random patch method and use this patch as an input to the model rather than the whole CXR image. Because the patch-based approach focuses more on the PICC lines compared to a model that takes the entire image as input, it can better preserve features of sparse signals such as PICC lines that may be lost due to the down sampling process within the model. In addition, it is useful noting that the patch-based approach can generate multiple patches from one image so provide sufficiently good performance even on a smaller training dataset \cite{zhao2018deep, coupe2011patch}. Similarly with these results, the proposed MFCN has also superior PICC line extraction and tip detection even with a small number of training samples, due to the patch-based characteristic of Stage 2 model.

Though Stage 2 effectively address the MFP problem caused by the sparseness of the PICC line and small training data, it is still difficult to completely solve the MFP issue through the patch-based approach of Stage 2 alone, if the PICC line is partially covered by anatomical structures or its edges are weakly expressed, Therefore, we proposed another network named PICC line reconnection network as Stage 3 that randomly creates a virtual disconnection on PICC line, allowing the network to perceive the disconnection extrinsically and complement it as a single PICC line. Through this method, it was possible to effectively improve the disconnection phenomenon caused by bone occlusion, thereby further improving the tip detection result.

In summary, the proposed MFCN has the following main contributions to overcome the limitations of existing PICC tip detection techniques: 

\begin{enumerate}
    \item (Multi-stage configuration to improve the performance of tip detection) By composing the proposed model in a multi-stage method, it can be combined with the existing conventional model to improve the performance of the existing model.
    \item (Second stage for optimizing PICC sparse line extraction) Through the patch-based segmentation technique, the network can focus more on the sparse area of PICC line in the entire CXR image, and the non-detection and false detection of this line are further improved.
    \item (Third stage for enhancing the line breakage robustness) By generating virtual multiple broken lines from the truth/whole PICC line and training the network to generate a complete PICC line from them, the network output can be robust to the MFP. Through this, it is possible to effectively improve the breakage caused by the overlapping of the PICC and other anatomical structures.
\end{enumerate}

We applied the proposed MFCN to the existing gold standard five DL-based models and experimentally evaluated further improvement in tip detection and PICC line extraction performance. Code is available at \href{https://github.com/kskim-phd/MFCN}{MFCN-link}.

\begin{figure}[t!]
    \centering
    \includegraphics[width=0.8\textwidth]{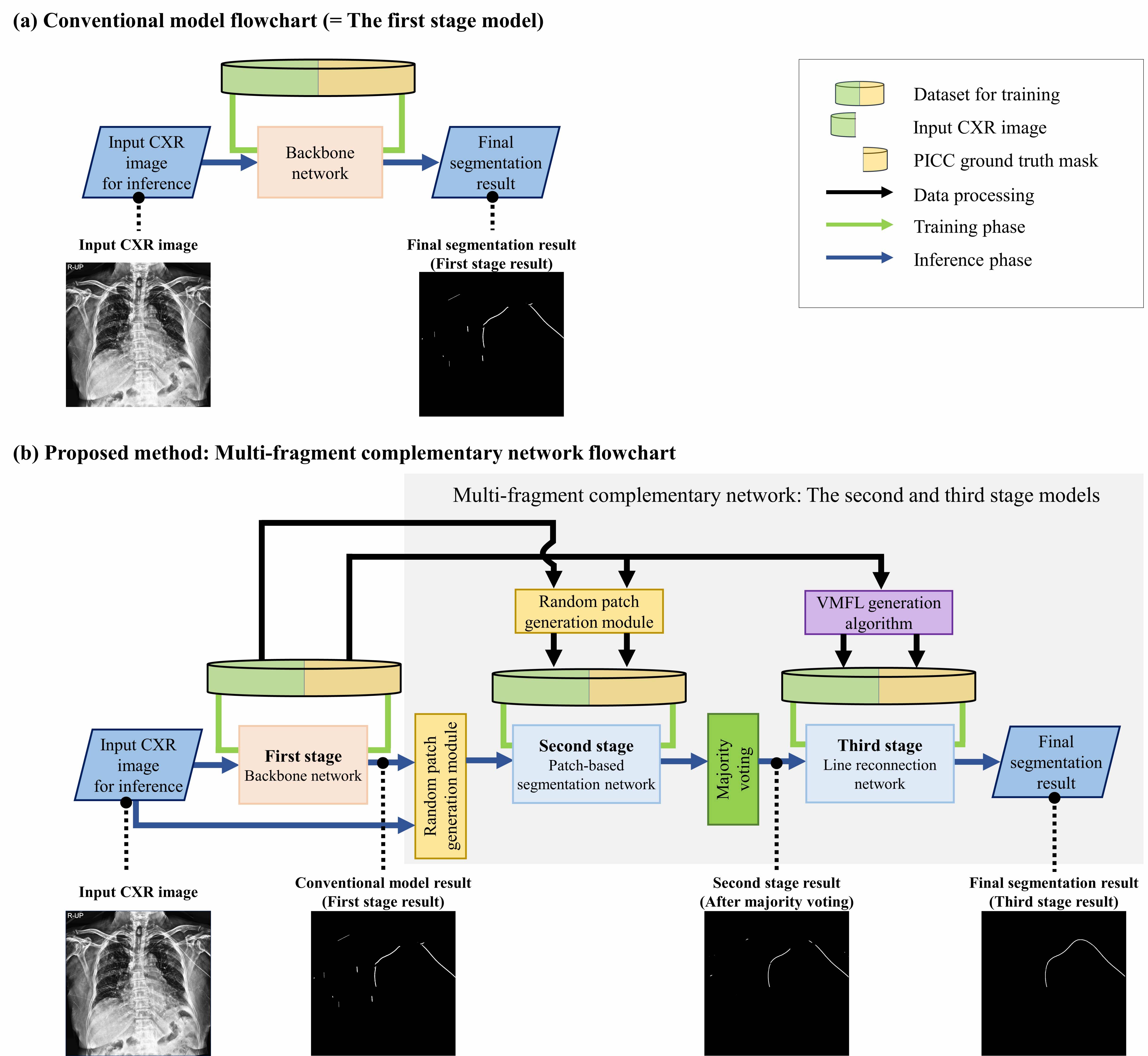}
    \caption{Schematic illustration of (a) baseline model and (b) proposed MFCN: MFCN consists of three stages where the supplementary process of second and third stages is added to the existing model (the first stage) to improve its PICC line restoration performance, thereby enabling precise tip detection.}
    \label{fig2:overview}
\end{figure}

\section{Methods}
\label{Sec:Methods}
\subsection{Overview of the proposed MFCN} \label{Sec:Overview}
We described the training and prediction process of the baseline method and the proposed MFCN in Figure \ref{fig2:overview}. The proposed MFCN (Figure \ref{fig2:overview}(b)) is a scheme designed to be additionally combined with any conventional PICC line segmentation network (Figure \ref{fig2:overview}(a)). Specifically, the proposed model consists of a total of three stages (Figure \ref{fig2:overview}(b)): the first stage is the conventional segmentation network (Figure \ref{fig2:overview}(a)), the second stage is the proposed patch-based segmentation network, and the third stage is the proposed line reconnection network. Both of the second and third stages are added to the first stage network to improve the PICC line segmentation performance and finally to detect the tip more accurately.  

In MFCN, the model for each stage is trained separately (illustrated as green line) with input and output data pre-processed (illustrated as black line) by some proposed external modules (e.g., random patch generation module in the second stage and virtual multi-fragment line generation (VMFL) module in the third stage). Specifically, the random patch generation module in the second stage generates a partial image by randomly cropping a patch containing PICC from the original CXR image in training data. Then the second stage trains the network to extract the partial PICC line corresponding to this generated patch by taking it as input. In the third stage, we generate a virtual multi-fragmented PICC line through the VMFL module and let the network to estimate its truth PICC line by taking it as input. 

\begin{figure}[t!]
    \centering
    \includegraphics[width=0.8\textwidth]{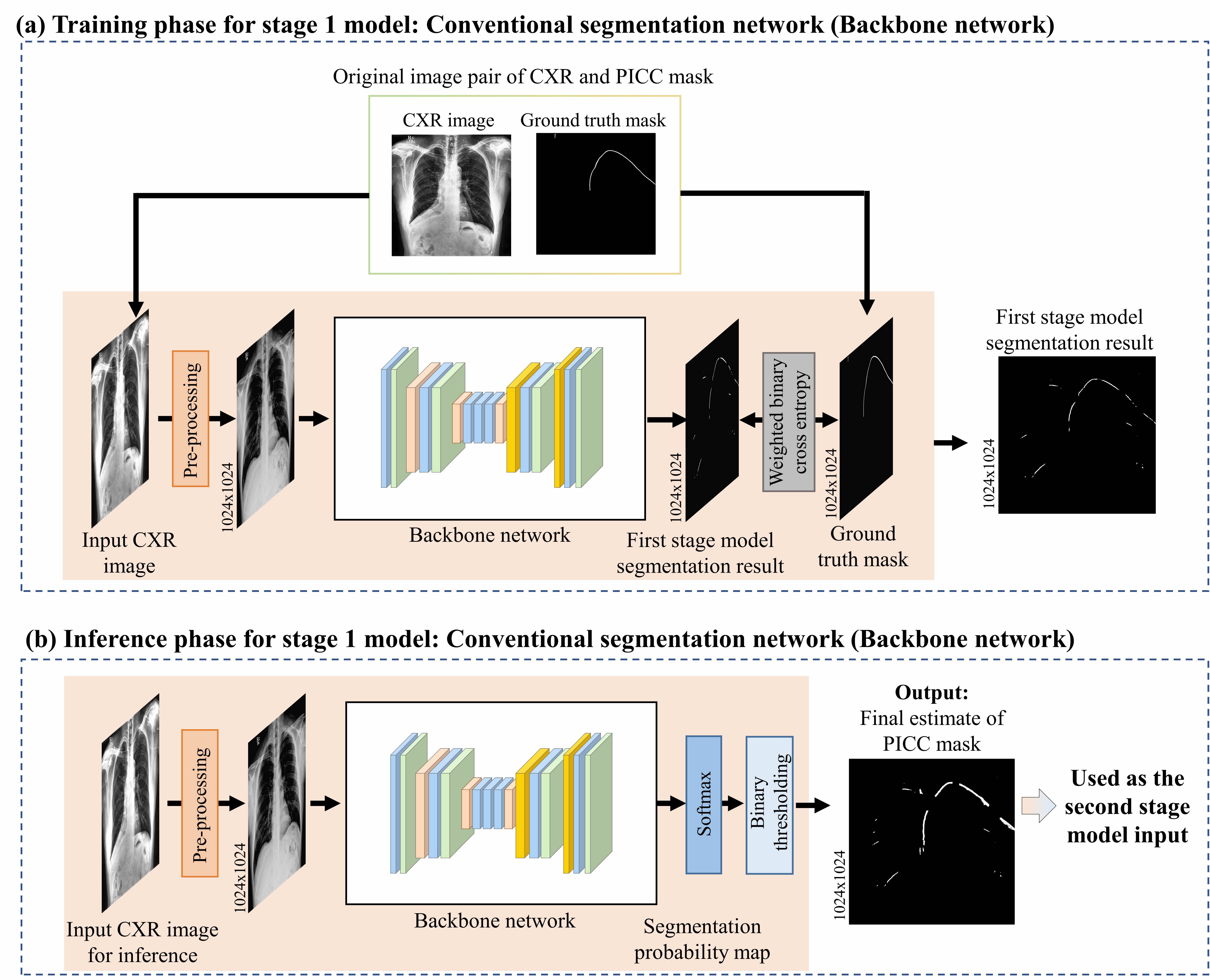}
    \caption{Schematic illustration of the first stage model: (a) the training and (b) inference phases}
    \label{fig3:first}
\end{figure}

In the inference phase, the trained model at each stage is combined together as the model output at the current stage is used as the input of that in the next stage as shown in the blue line in Figure \ref{fig2:overview}(b). Specifically, in the case of the second stage, segmentation result for the partial PICC line given by each patch output is aggregated as the complete prediction result of the whole PICC line through majority voting. And this prediction result for the whole PICC line is used as the input of the third stage model. As a result, as the CXR image goes through the stages of the inference phase, it can be observed from the result image of each stage at the bottom of Figure \ref{fig2:overview} that the complete PICC line is gradually extracted without any break or interruption.

It is useful to note that the architectures of the proposed second and third stage models do not have any specific structure and can be applied to any architecture of the existing networks for segmentation (e.g., FCN \cite{long2015fully}, U-Net (UNET) \cite{ronneberger2015u}, FC-DenseNet (FCDN) \cite{jegou2017one}, Attention U-Net (AUNET) \cite{oktay2018attention}). For the architectures of the proposed second and third stage models, we selected FCDN among existing segmentation networks, as it performs better or similar compared to the other models. We detailed it in section \ref{Sec:backbone}.

\subsection{The first stage network}

The first stage network is a model that receives a whole CXR image as an input and provides a binarized map of the same size as an output, consisting of 1 for pixels belonging to PICC and 0 for pixels not belonging to PICC. The first stage model can be set as any conventional PICC line segmentation network like the baseline model as shown in Figure \ref{fig2:overview}(a). The goal of the proposed MCFN is to improve the first stage model’s segmentation performance by adding the second and third stage models. In the experimental section, we set up various existing models as this first stage network and proved that the proposed MFCN consistently improved performance for each model.

\subsubsection{Training phase for the first stage network}

Figure \ref{fig3:first}(a) depicts the learning process of the first stage model. The original CXR image was resized to have 1024$\times$1024 resolution, then the contrast-limited adaptive histogram equalization (CLAHE) \cite{pizer1990contrast, reza2004realization} was performed as pre-processing so that the PICC could be seen more clearly. After that, the pre-processed image was used as an input of the first stage backbone network, and the network was trained to let the segmentation result of PICC line have a pixel value of 1 and the other areas have a pixel value of 0. Since individual pixel output values of the model were expressed as probability values between 0 and 1, we followed the conventional approach by extracting the PICC line through minimizing (the weighted) cross-entropy loss.

\begin{figure}[t!]
    \centering
    \includegraphics[width=0.8\textwidth]{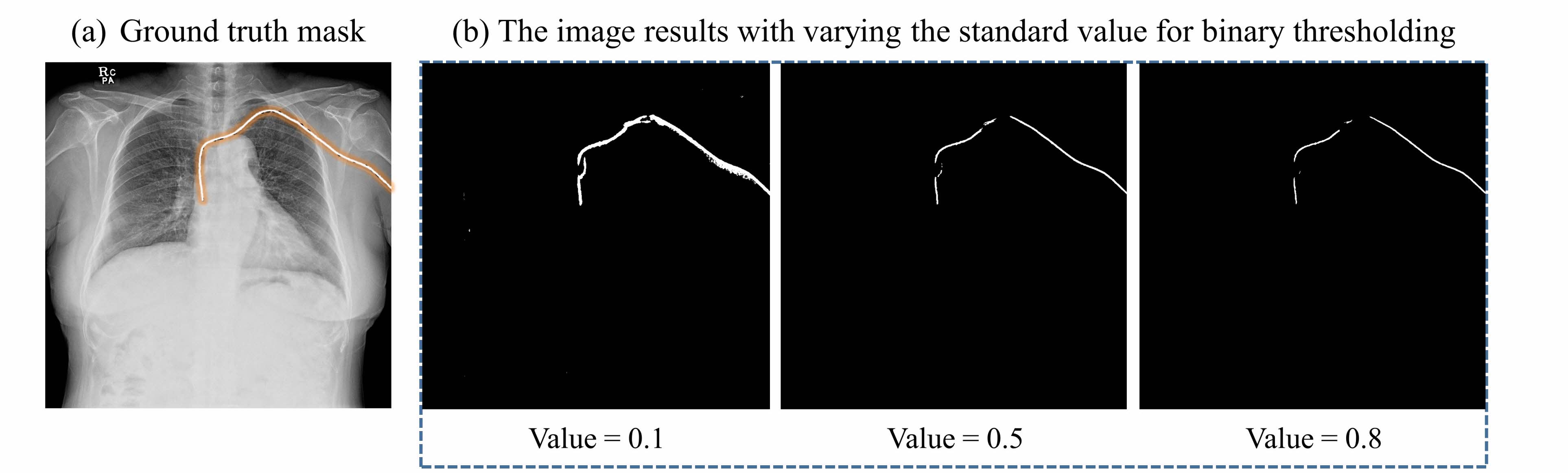}
    \caption{Description of various examples with varying thresholds for the final PICC line prediction results of the first stage network}
    \label{fig4:thresh}
\end{figure}

\subsubsection{Inference phase for the first stage network}

In the inference phase of the first stage network, different from the training phase, more relaxed thresholding is given to allow false-positive (FP) appropriately, but set to reduce false-negative (FN). The rest of the process proceeds in the same process as the training process to obtain the model output. Note that each pixel in the first stage probability map (i.e., after the softmax layer) is distributed between 0 and 1. In this study, we binarized it based on a threshold of 0.01 to make the final PICC mask. While most existing works select 0.5 as this threshold value, we set it to 0.01 to further reduce FN even if some FP of the PICC line is allowed.

As shown in Figure \ref{fig4:thresh}, the PICC prediction result obtained through a low thresholding value is thicker and has less breakage. However, lowering the threshold also cause FP like sporadic noise and in particular it still does not fundamentally solve the MFP, which is further resolved through next stages.

\subsection{The second stage network: Patch-wise PICC segmentation network}

The proposed second network corrects the prediction result of the PICC line of the first stage network through patch-based PICC segmentation. The training and testing methods were covered in Sections \ref{Sec:patch_train} and \ref{Sec:patch_inference} respectively, and prior to these, in Section \ref{Sec:patch_generation}, the method of generating patch data is introduced. As the second stage network learns by extracting multiple local patch images focused on the PICC from a single whole CXR image, it increases the relative PICC area within the input image of the network with preserving the resolution of the original CXR image. These make the network more aware of the PICC area without degradation such as by downsampling, thereby enhancing the MFP problem. 

\subsubsection{Data pre-processing: Random patch generation}
\label{Sec:patch_generation}
The main difference of the proposed second stage model to the first stage one is that it does not take as input the entire CXR image but rather a patch composed of its partial region. The patch generation process was performed differently in the training and testing process of the model as follows. Each patch was made to have a size of 512$\times$512, while the original CXR image, which the patch is extracted from, has the size of 2017$\times$2017-3408$\times$3040 (Table \ref{Table1:dataset}).

\begin{figure}[t!]
    \centering
    \includegraphics[width=0.8\textwidth]{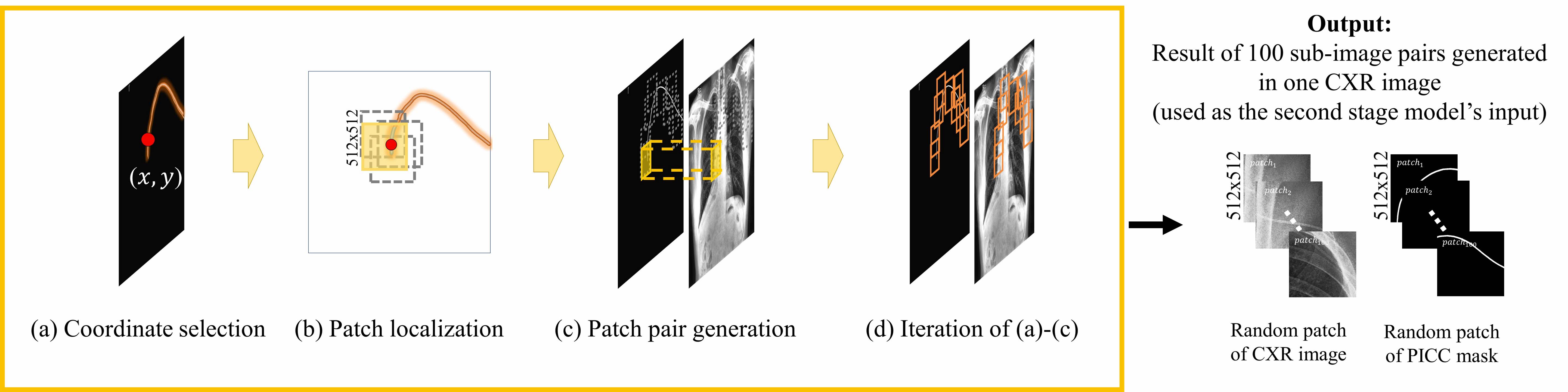}
    \caption{Illustration of the patch generation module in the training phase: (a) Randomly selecting a two-dimensional coordinate in the PICC area, (b) Generating a patch by cropping the CXR image to include the coordinate, (c) Creating a pair of CXR and PICC mask sub-images in the same section as the patch and use it as an input and output for model training, (d) Repeat the process of (a)-(c) 100 times}
    \label{fig5:patch}
\end{figure}

For training phase, we were given the truth binary masks of PICC line for CXR training data. Within each CXR individual image of the training data, we randomly selected a two-dimensional coordinate in the truth PICC area and randomly generated a patch so that it could contain this coordinate at a random location within the patch. In this way, 100 patches were generated from each sample in training data, therefore the data was augmented 100 times. We cropped the original CXR and PICC mask images at the same location to generate 100 patch pairs and used them as input and output for model training. We presented the overall process of generating these pairs in Figure \ref{fig5:patch}.

For inference phase, we were given the binary mask prediction results of PICC line for CXR testing data as the first stage model’s outputs. We then randomly selected a two-dimensional coordinate in the predicted PICC line area (i.e., the positive region of pixel value equal to 1 in this binary mask prediction) and randomly generated a patch for it to include this coordinate at a random location. We randomly produced 200 patches from each CXR test image through this way and let them to be taken individually as input of the second stage model.

\begin{figure}[t!]
    \centering
    \includegraphics[width=0.8\textwidth]{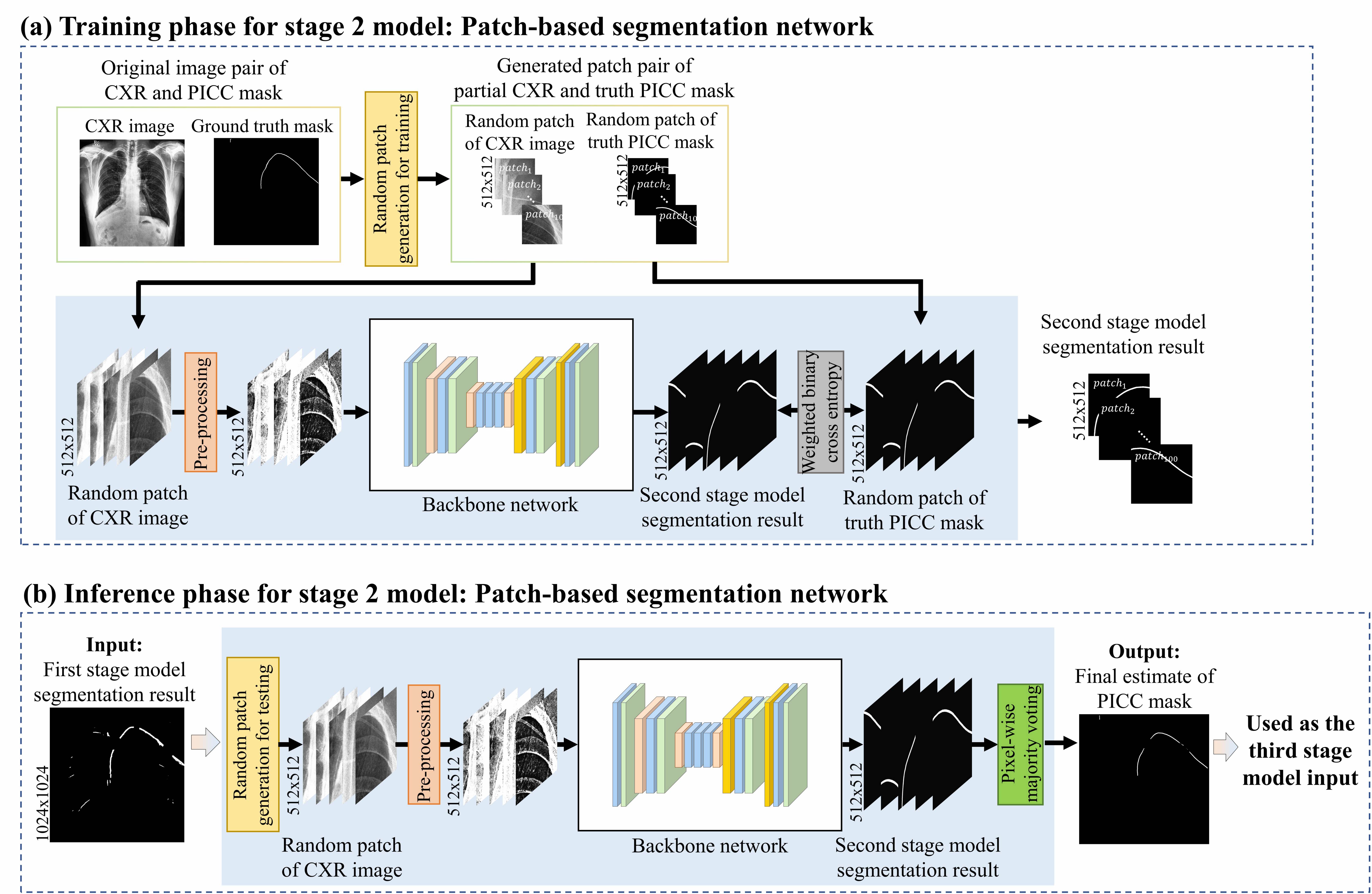}
    \caption{Schematic illustration of the second stage model: (a) the training and (b) inference phases}
    \label{fig6:secon}
\end{figure}

\subsubsection{Training phase for the second stage network}
\label{Sec:patch_train}

Figure \ref{fig6:secon}(a) depicts the learning process of the second stage model. For each sample (i.e., patient) in training data, 100 patch pairs were generated as detailed in Section \ref{Sec:patch_generation}. Each patch pair consists of the CXR partial image and partial mask of truth PICC corresponding to the same area and they were taken as the input and output label of the second stage network, thereby learning the network to correctly estimate the PICC partial mask as its output. For each patch, CLAHE was applied as a preprocessing to improve the image recognition rate of the network. Though any segmentation network can be applied as a backbone of the second stage model, for simplicity, we adopted FCDN proposed by Jegou \textit{et al}. \cite{jegou2017one}. Each pixel value of the model output is expressed as a probability value between 0 and 1, and the binary cross entropy loss is applied in learning the network to output 1 for the target PICC region. 

\subsubsection{Inference phase for the second stage network}
\label{Sec:patch_inference}

The inference phase of the second stage model is illustrated in Figure \ref{fig6:secon}(b), which consists of a process similar to the training phase but mainly differs in the following two aspects: 1) The input patch was generated based on the prediction result of the PICC mask (i.e., the output of the first stage model), not the PICC truth mask, 2) An additional process is required to create a complete PICC segmentation mask by converting all prediction results of patch unit into one prediction result of the whole image unit. We performed it by developing a pixel-wise majority voting approach as shown in the green box in Figure \ref{fig6:secon}(b).

\begin{figure}[t!]
    \centering
    \includegraphics[width=0.8\textwidth]{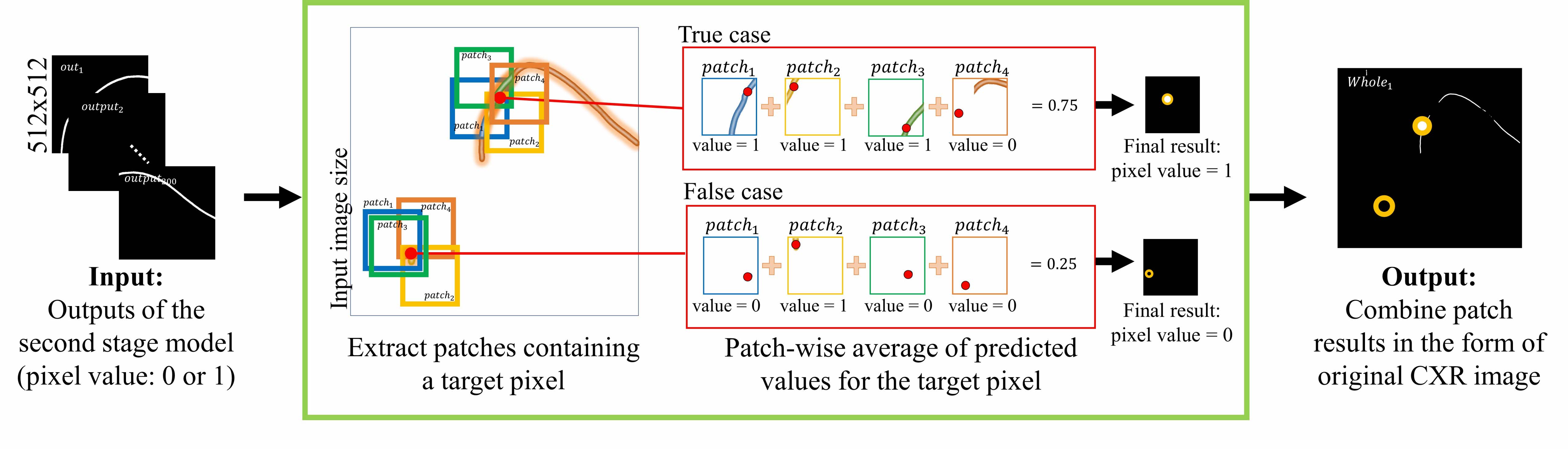}
    \caption{Illustration of the pixel-wise majority voting scheme: Suppose that there exist 4 output patches of the second stage model including a specific coordinate on the whole CXR image, the pixel values of the corresponding coordinate are averaged, binarized via a hard thresholding, and assigned to the corresponding coordinate. Through this process, the entire PICC line is predicted by reconstructing individual patches into one whole image.}
    \label{fig7:majority}
\end{figure}

Specifically, the second stage model is designed to extract the PICC line corresponding to the input patch region, which implies that the individual patch output of the model focus only on a specific region of interest in the entire CXR image. The proposed pixel-wise majority voting approach is the post-processing of the model’s output patches to produce a whole PICC segmentation/prediction mask by combining all prediction results of patches for some partial regions of the PICC mask. We illustrated its detailed process in Figure \ref{fig7:majority}. Given 200 patches of the model’s outputs for partial PICC line estimates, the pixel-wise majority voting approach mainly consists of the following three steps: (a) Select a specific coordinate on whole CXR image, (b) Calculate the average value of the selected coordinate for all patches including the corresponding coordinate. (c) Assign a binary value at the selected coordinate as a result of applying the binary thresholding to this average value, (d) Repeat the process of (a), (b), and (c) for every coordinate on whole CXR image. We set the threshold to 0.7 to produce a final prediction result of the PICC mask (as a binary map with the same size as the whole CXR image), with a value of 1 (positive), if the corresponding average value is greater than or equal to the threshold, and 0 (negative) otherwise.

In Figure \ref{fig7:majority}, we exemplified a case where only 4 patches contain a certain coordinate. Given three patches predict the corresponding coordinate value as positive (i.e., 1) but one patch predicts false (i.e., 0), the average result is given as 0.75. As it is higher than our threshold 0.7, so we finally determined the corresponding pixel value as positive (i.e., 1). Though we took a simple example of 4 patches in Figure \ref{fig7:majority}, 200 patches are created in fact so that multiple patches sufficiently overlap for each coordinate, helping to filter the errors of minor patches more stably. Therefore, this result indicates that 25\% of patch errors are filtered from the proposed pixel-wise voting approach, making PICC segmentation result and its MFP improved more than those of the first stage network.  

\subsection{The third stage network: Line reconnection network}
The second network result effectively improves the MFP problem, but does not completely solve the problem of micro-breaking of the PICC line due to the occlusion of the bone and the PICC line. To solve this, we created a virtual break of the PICC line and let the  third stage network recognize it. We first introduce the process of generating a virtual break in Section \ref{Sec:VMFLG} and then the learning/testing phase of the network in Sections \ref{Sec:third_train} and \ref{Sec:third_inference}.

\subsubsection{Data pre-processing: Virtual multi-fragment line generation (VMFLG)}
\label{Sec:VMFLG}

\begin{figure}[htb!]
    \centering
    \includegraphics[width=0.8\textwidth]{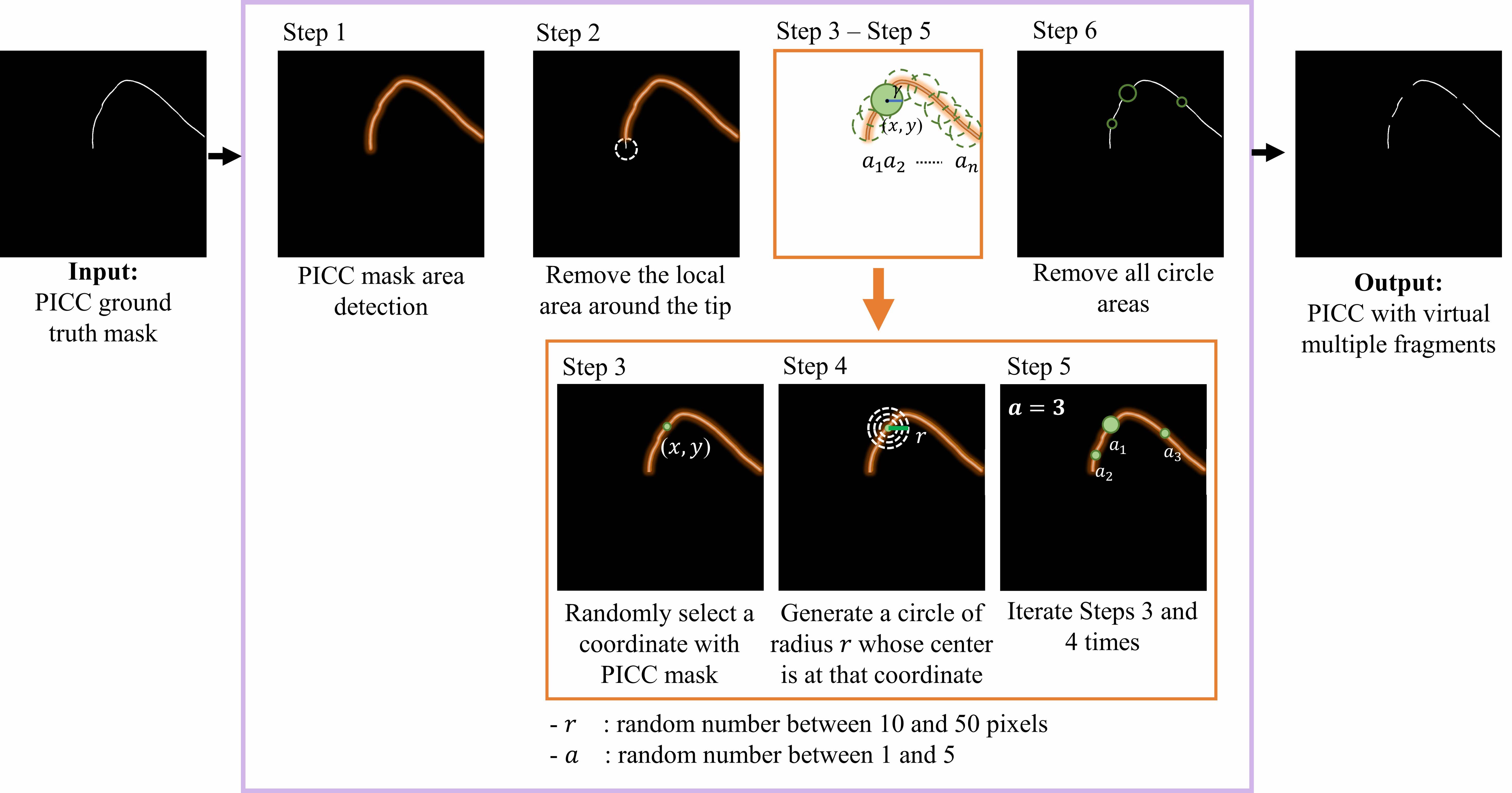}
    \caption{Virtual multi-fragment line generation (VMFLG) algorithm}
    \label{fig8:VMFLG}
\end{figure}

Figure \ref{fig8:VMFLG} introduces the overall process to create a virtual multi-fragment line of PICC. We segment the PICC mask in Step 1, select a two-dimensional ($x$,$y$)-coordinate in Step 3 in this PICC area, remove a circle area of radius randomly ranged from 10 to 50 pixels around this coordinate to make a breakpoint in PICC in Steps 3-4, repeat this removal process (Step 5), and finally generated the virtual multi-fragment line of PICC. We generate these virtual lines by 10 for every PICC image for training the network. We segment the PICC mask in Step 1 and specify where the breakpoint will be made in Step 2. We designed this virtual PICC line to maintain its original tip (Step 2). As it makes the network explicitly aware of the tip location, the network does not predict further extension of the PICC line beyond the tip. 

\subsubsection{Training phase for the third stage network}
\label{Sec:third_train}

We learned the third stage model by taking as its input data of the virtual multi-fragment PICC lines generated by the VMFLG and let it predict the actual PICC line as its output by minimizing the pixel-wise cross entropy loss between the PICC ground truth mask and the model output. It is useful to note that as stage 1 and stage 2 models use CXR data as input but the corresponding stage 3 model uses binary image as input, we did not proceed with any separate preprocessing such as CLAHE in the stage 3 model.

\begin{figure}[htb!]
    \centering
    \includegraphics[width=0.8\textwidth]{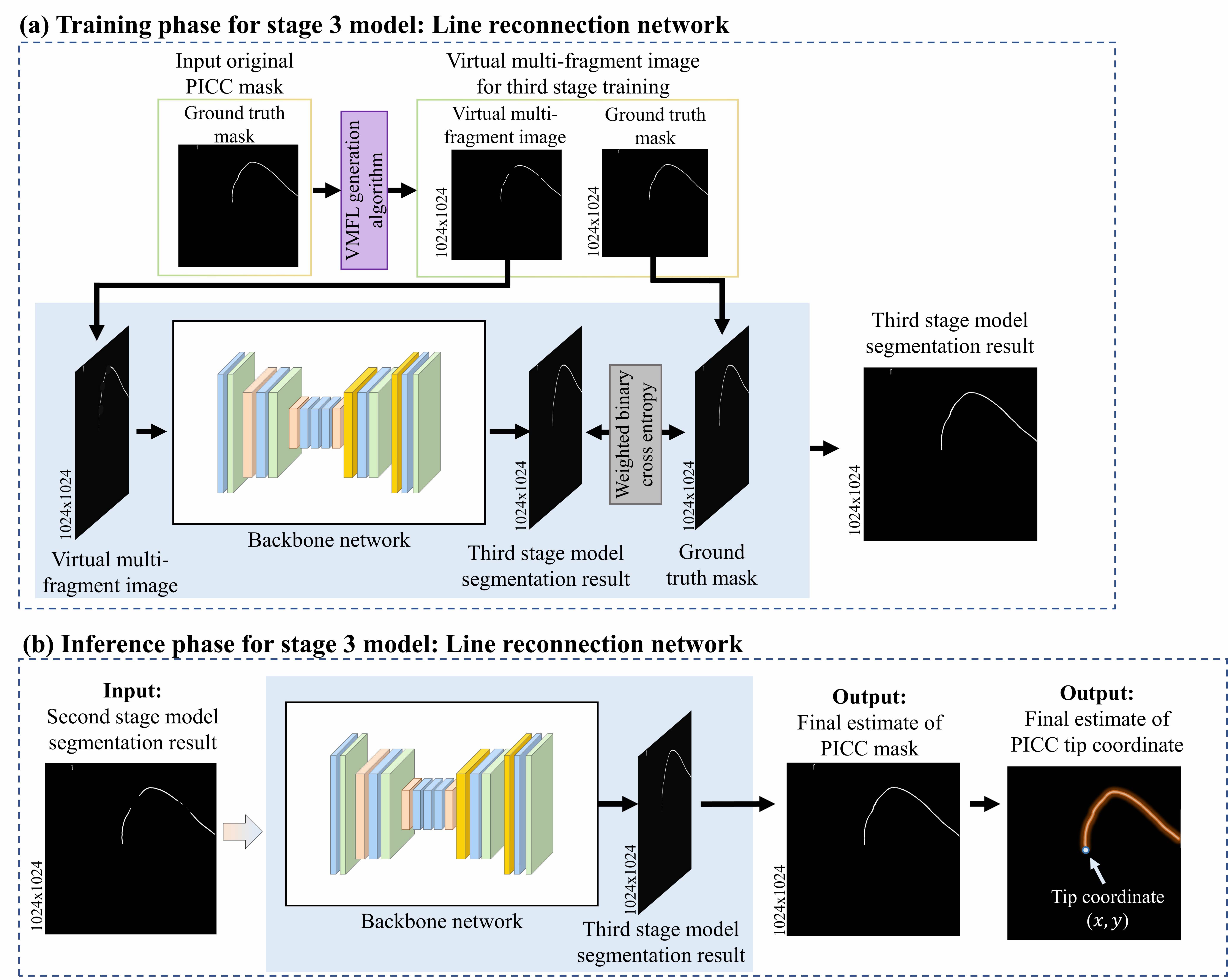}
    \caption{Schematic illustration of the third stage model: (a) the training and (b) inference phase}
    \label{fig9:third}
\end{figure}

\subsubsection{Inference phase for the third stage network}
\label{Sec:third_inference}

As input of the stage 3 network pretrained as in Section \ref{Sec:third_train}, we took the PICC line segmentation result of the stage 2 model. Then, we got the final segmentation result of the PICC line from its output. This process is illustrated in Figure \ref{fig9:third}(b). 

Through this three-stage refining operation, a more complete PICC line can be estimated than the existing AI methods (i.e., the output result of the first stage network). We tracked down from the top of the PICC prediction line to the bottom until there existed a breakpoint and the final tip location is obtained by selecting the coordinate with this breakpoint.

 \section{Results}
\label{Sec:Results}

\subsection{Dataset}

\subsubsection{Data collection and setup for internal validation data}

CXR images of patients from 2017-01-01 to 2020-12-31 were collected as a format of digital imaging and communications in medicine (DICOM) from X-ray devices (Manufacturer: GE Healthcare, Samsung Electronics) and they all were anonymized. All collected CXR images were posterior-anterior (PA) type, and their size is ranged from 2017$\times$2017 to 3408$\times$3040 pixels, but most of the images were composed of 2021×2021. Among them, we did not include cases in which a CVC other than PICC was inserted or only a part of the chest was visible. The study design was approved by the institutional review board of our institution (approval number: 2021-05-164). The requirement for informed consent was waived owing to the retrospective nature of this study.

A total of 280 images collected were divided into 130 as data for learning and 150 as data for testing (Table \ref{Table1:dataset}). In the process of dividing, the number of cases where the PICC was inserted from the right side or was misplaced were randomly divided without fixing them at a certain ratio. As a result, in the training (testing) set, 15 and 7 (22 and 13) samples were the right-side PICC insertion and misposition cases. Among the training dataset, 100 samples were used for training the network and 30 were used as validation data. Ground truth of PICC line mask for collected data was annotated by board certified radiologists as a form of binary image (foreground pixel=1, background pixel=0). For this annotation, they used the functions of drawing essential region of international (brush and change) supported by OsiriX (open-source software; www.osirixviewer.com) \cite{rosset2004osirix, rosset2005general}. Then, the PICC ground truth masks and their PICC tip endpoint coordinates have passed follow-up review by board certified radiologists.  

\begin{table}[hbt!]
\footnotesize
\centering
\resizebox{0.7\linewidth}{!}{%
\begin{tabular}{ccc}
\hline
 & \begin{tabular}[c]{@{}l@{}}Learning set (\#130)\end{tabular} & Test set (\#150) \\ \hline
Minimum image size (pixel) & \begin{tabular}[c]{@{}c@{}}2021 × 2021\end{tabular} &  \begin{tabular}[c]{@{}c@{}}2017 × 2017\end{tabular} \\
Maximum image size (pixel) & \begin{tabular}[c]{@{}c@{}}3408 × 3040\end{tabular} &  \begin{tabular}[c]{@{}c@{}}3040 × 2902\end{tabular} \\
View point & Posterior-Anterior &  Posterior-Anterior \\
Pixel spacing (mm) & 0.125, 0.14, 0.143, 0.194 &  0.14, 0.143, 0.194 \\
\begin{tabular}[c]{@{}l@{}}Right insertion case (\#)\end{tabular} & 15 &  22 \\
Misposition case (\#) & 7 &  13 \\ \hline
\end{tabular}%
}
\caption{Characteristics of training and test set of CXR data for internal validation}
\label{Table1:dataset}
\end{table}

\begin{figure}[htb!]
    \centering
    \includegraphics[width=0.8\textwidth]{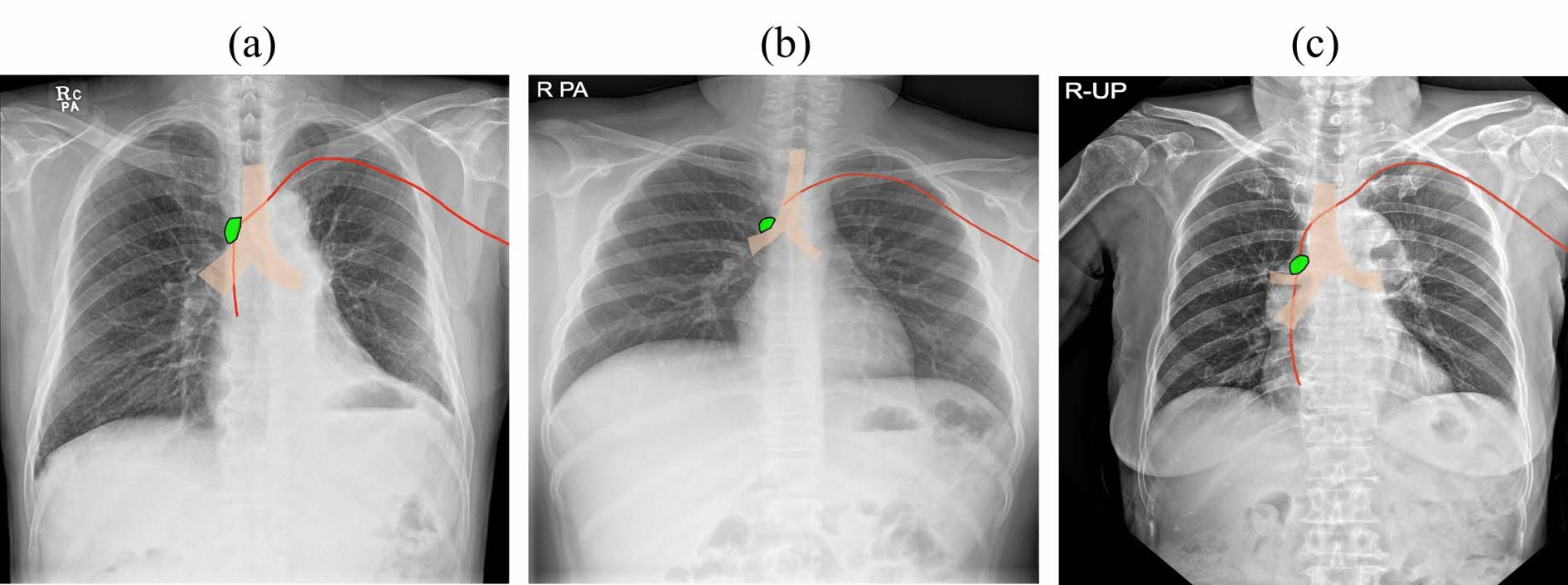}
    \caption{CXR images of collected dataset for PICC insertion: (a) an ideally located PICC catheter case, (b) mispositioned PICC catheter case - tip located too high, (c) mispositioned PICC catheter case - tip located too low, (red area) ground truth PICC mask, trachea (orange area), azygos vein confluence (blue area)}
    \label{fig10:CXR}
\end{figure}

An ideal location of the PICC is defined as being located at the junction of the superior vena cava and the right atrium which is the level of the lower aspect of the bronchus intermedius (Figure \ref{fig10:CXR}) \cite{li2018randomized}. In our study, we defined the misposition of PICC as the case where the PICC tip endpoint was located above the azygos vein confluence or in the right atrium or ventricle level (Figure \ref{fig10:CXR}). 

\subsubsection{Data collection and setup for external validation data} \label{Sec:external_data}

For external validation, we used the publicly available dataset of the Royal Australian and New Zealand college of radiologists (RANZCR). This dataset consists of more than 40,000 CXRs among ChestXRAY14 dataset provided by the national institutes of health \cite{wang2017chestx}. The RANZCR is a not-for-profit professional organization for clinical radiologists and radiation oncologists in Australia, New Zealand, and Singapore. The RANZCR dataset consists of endotracheal tube, nasogastric tube, central venous catheter (CVC), and Swan-Ganz catheter, and is divided into abnormal, borderline, and normal depending on the location. For the verification of baseline and proposed models, 280 data were randomly selected from normal CVC images including PICC (1200 images).
Among them, 100, 30, and 150 data were used as training, validation (for hyperparameter tuning), test datasets respectively. The image size of the collected data was ranged from 1610×1734 to 3056×3056 and the case where PICC was inserted from the right (left) side was distributed to 44\% (56\%) in the training and validation dataset and 50\% (50\%) in the test dataset. 

\subsubsection{Data preprocessing}

As the collected CXR images have characteristics of low-pixel contrast and high-noise \cite{sandborg2006comparison, baath2005nodule}, we performed the following two pre-processing approaches to normalize their contrast and dimension. 

The first pre-processing is to increase image contrast of the input image by applying the contrast limited adaptive histogram equalization (CLAHE) \cite{pizer1990contrast, reza2004realization}. CLAHE has been widely applied to many medical images as it can improve contrast and reduce noise implication. In this study, we used CLAHE as the pre-processing step of each model using the entire CXR image (e.g., the first stage model) or the model using the patch image of CXR image (e.g., the second stage model). Specifically, we applied CLAHE to the entire CXR image and in the first stage model each patch after generating multiple patches from one entire CXR image in the second stage model. 

The second pre-processing is to resize the original CXR image. Note that the collected CXR image consists of various image sizes from 2017$\times$2017 to 3408$\times$3040 pixels (Table \ref{Table1:dataset}) with an average size of 2021$\times$2021 pixels. The first and third stage models received up to images of size 1024$\times$1024 as input due to GPU memory problems, so we resized the input image size at 1024×1024 in the first and third stage models. In the second stage model, the patch of size 512×512 was extracted from the original CXR image without using any resize on the original image in order to preserve it as it is. 

\begin{figure}[htb!]
    \centering
    \includegraphics[width=0.8\textwidth]{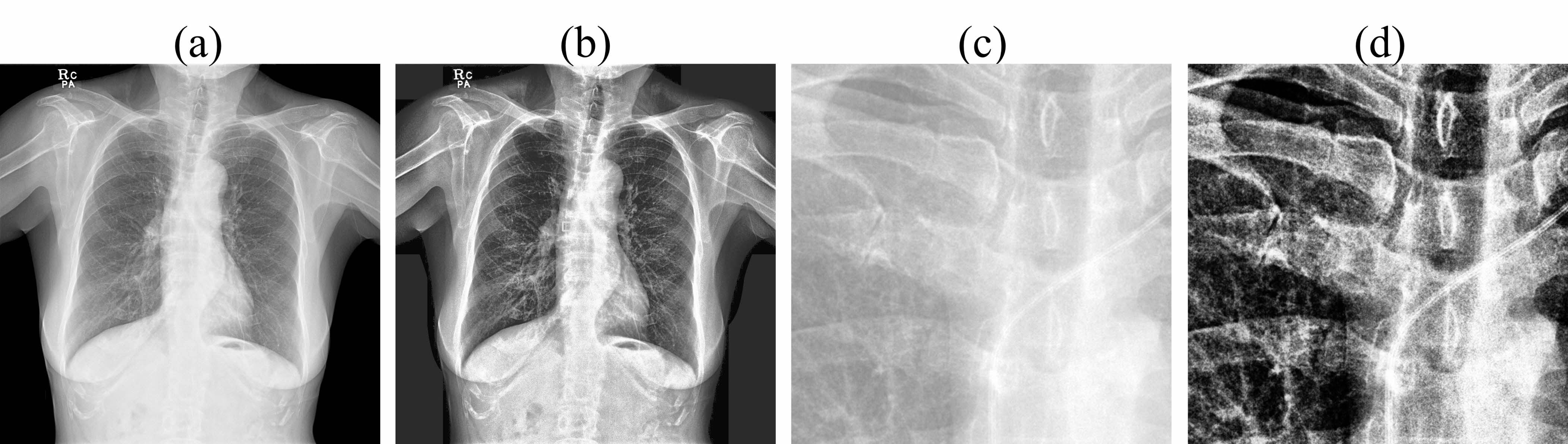}
    \caption{Illustration of application with CLAHE: the original CXR image (a) without or (b) with CLAHE and the patch of CXR image (c) without or (d) with CLAHE.}
    \label{fig11:CLAHE}
\end{figure}

\subsection{Implementation detail}
\subsubsection{Setup for conventional method and proposed MFCN}

MFCN was applied to five conventional models to verify the effect of improving their performances by MFCN. These conventional models consist of three representative DL-based models developed to solve the general semantic segmentation problem (i.e., FC-DenseNet (FCDN) \cite{jegou2017one}, UNET \cite{ronneberger2015u}, and Attention U-Net (AUNET) \cite{oktay2018attention}) and two latest DL-based models developed to detect PICC tip more precisely (FC-DenseNet with hough transform (FCDN-HT) \cite{lee2018deep} and UNET with Region Proposal Network (UNET-RPN) \cite{yu2020detection}). FCDN-HT is a model in which the probabilistic hough transform \cite{duda1972use, kiryati1991probabilistic} provided by OpenCV is added to the FCDN model result as a post-processing process. Though authors of FCDN-HT used a naive fully convolutional network (FCN) rather than its dense adaptation (FCDN), as they suggested that updating FCN with a dense architecture (e.g., FCDN) can further improve tip detection accuracy \cite{lee2018deep, wang2019dense}, we adopted FCDN instead of FCN in reproducing their work. We also used the HT parameter values as suggested by the authors \cite{lee2018deep}. UNET-RPN is a model that detects the location of the PICC tip by additionally combining the RPN module of Faster R-CNN \cite{ren2015faster} with UNET. 

We set the above five existing individual models as the first stage model of MFCN, and for simplicity, we applied FCDN for both second and third stage models of MFCN. Accordingly, the validity of the proposed technique was verified by determining whether there was a performance improvement when the second/third stage model was added (i.e., proposed scheme) compared to when only the first stage model was used (i.e., basic scheme). It is useful to note that the base network of the second and third models in MFCN can be set as any existing segmentation model. The detailed performance comparison was presented in Section \ref{Sec:backbone}.

\subsubsection{Hyperparameter setup for proposed MFCN}

The proposed MFCN was trained for 100 epochs for each experiment with initial learning weight set to 0, the batch size was set to 2, and Adam optimizer \cite{kingma2014adam} with a learning rate of $1e^{-4}$ was used. In order to prevent overfitting, the dropout rate \cite{srivastava2014dropout} was set to 0.2 and early stopping was performed based on the validation loss. Weighted binary cross entropy was used as a loss function for model training and validation, and the balance weight was set to 0.5 in background. All experiments were run on two V100 graphics processing units (GPUs) and Pytorch (1.4.0) under Python 3.6.

\subsection{Evaluation metrics}

PICC segmentation accuracy and tip detection accuracy were measured for quantitative evaluation of the proposed MFCN. The PICC segmentation result is evaluated by calculating the Dice similarity coefficient (DSC) between the ground truth PICC mask and the model segmentation result as formulated as follows: 
\begin{equation}\label{eqn:DSC}
    \textup{DSC} = {\frac{2|\textup{GT} \cap \textup{PD}|}{|\textup{GT}| + |\textup{PD}|}}
\end{equation}

where $\textup{GT}$ is the pixel region of PICC ground truth mask and $\textup{PD}$ is that of its model prediction. Tip detection accuracy is evaluated using root mean square error (RMSE) to evaluate the distance between ground truth tip position coordinates and predicted tip position coordinates. The RMSE was calculated as follows: 
\begin{equation}\label{eqn:RMSE}
    \textup{RMSE} = \sqrt{\sum\limits_{i=1}^n \frac{(x_{pred} - x_{gt})^2+(y_{pred} - y_{gt})^2}{n}}
\end{equation}

where $x_{gt}$ and $y_{gt}$ are $x$ and $y$ coordinates of the ground truth tip, $x_{pred}$ and $y_{pred}$ are $x$ and $y$ coordinates of the predicted tip, and $n$ denotes the total number of test images.

\subsection{Internal validation result} \label{Sec:internal valid}
\subsubsection{PICC tip location detection}

We presented the performance comparison results between the proposed MFCN and the existing technologies for PICC tip location detection in Table \ref{Table2:inter_r}. We presented in the first column the mean and variance of RMSE, in the second column the ratio (\%) of test samples for the predicted PICC segmentation line result to be derived as one complete line (correctly estimated), and in the third column the ratio (\%) of test samples when their RMSE is sufficiently low (i.e., below 1cm). As a result, the existing results showed an average RMSE of 20 mm or more, whereas the MFCN showed an average RMSE of 10 mm or less, thereby consistently improving the results of each of the existing models by reducing their RMSE by less than half (the first column in Table \ref{Table2:inter_r}). The ratio of the number of reconstructed single complete PICC lines without MFP was less than 50\% on average in the existing model, but when the proposed method was applied, it was more than 90\%. Therefore, the proposed MFCN has the MFP problem improvement rate by more than 40\% (the second column in Table \ref{Table2:inter_r}). In addition, the proposed MFCN increased the precision/correct detection of PICC tip location (i.e., RMSE is less than 1 cm) by at least 10\% (the third column in Table \ref{Table2:inter_r}).

\begin{table}[hbt!]
\footnotesize
\centering
\resizebox{0.7\textwidth}{!}{%
\begin{tabular}{ccccccc}
\hline
 & \multicolumn{2}{c}{RMSE (mean±sd, mm)} & \multicolumn{2}{c}{No MFP (\%)} & \multicolumn{2}{c}{$|RMSE|$ \textless 1cm (\%)} \\ \cline{2-7} 
\multirow{-2}{*}{\begin{tabular}[c]{@{}c@{}}Conventional\\ Model name\end{tabular}} & Baseline & \textbf{MFCN} & Baseline & \textbf{MFCN} & Baseline & \textbf{MFCN} \\ \hline
FCDN \cite{jegou2017one} & 23.37 ± 32.12 & \textbf{9.37 ± 18.13} & 49 & \textbf{93} & 60 & \textbf{80} \\
UNET \cite{ronneberger2015u} & 34.98 ± 39.93 & \textbf{9.11 ± 16.23} & 47 & \textbf{89} & 48 & \textbf{81} \\
AUNET \cite{oktay2018attention} & 35.80 ± 41.52 & \textbf{10.98 ± 21.05} & 35 & \textbf{84} & 47 & \textbf{78} \\
FCDN-HT \cite{lee2018deep} & 17.16 ± 28.25 & \textbf{9.36 ± 18.22} & 62 & \textbf{90} & 71 & \textbf{81} \\
UNET-RPN \cite{yu2020detection} & 22.93 ± 16.46 & \textbf{9.77 ± 18.87} & 25 & \textbf{87} & 26 & \textbf{81} \\ \hline
Average & 26.85 & \textbf{9.72} & 43.6 & \textbf{88.6} & 50.4 & \textbf{80.2} \\ \hline
\end{tabular}%
}
\caption{Internal validation result for PICC tip location detection: Comparison of RMSE results between the existing methods and those applied with the proposed MFCN}
\label{Table2:inter_r}
\end{table}

\subsubsection{PICC line segmentation}

We also presented the PICC line segmentation results in Table \ref{Table3:inter_s}. When the proposed MFCN was applied to the existing method, Dice score performance was improved in all cases, and it was confirmed that the sample probability of extracting the PICC line with high accuracy of Dice score 0.95 or higher was at least 10\% higher on average compared to the existing method.

\begin{table}[hbt!]
\footnotesize
\centering
\resizebox{0.7\textwidth}{!}{%
\begin{tabular}{ccccc}
\hline
 & \multicolumn{2}{c}{Dice score (mean±sd, mm)} & \multicolumn{2}{c}{Dice score \textgreater 0.95 (\%)} \\ \cline{2-5} 
\multirow{-2}{*}{\begin{tabular}[c]{@{}c@{}}Conventional\\ Model name\end{tabular}} & Baseline & \textbf{MFCN} & Baseline & \textbf{MFCN} \\ \hline
FCDN \cite{jegou2017one} & 0.91 ± 0.16 & \textbf{0.92 ± 0.17} & 60 & \textbf{76} \\
UNET \cite{ronneberger2015u} & 0.90 ± 0.16 & \textbf{0.92 ± 0.17} & 59 & \textbf{75} \\
AUNET \cite{oktay2018attention} & 0.90 ± 0.16 & \textbf{0.92 ± 0.17} & 55 & \textbf{75} \\
FCDN-HT \cite{lee2018deep} & 0.90 ± 0.14 & \textbf{0.92 ± 0.17} & 47 & \textbf{79} \\
UNET-RPN \cite{yu2020detection} & 0.91 ± 0.12 & \textbf{0.92 ± 0.17} & 52 & \textbf{76} \\ \hline
Average & 0.904 & \textbf{0.92} & 54.6 & \textbf{76.2} \\ \hline
\end{tabular}%
}
\caption{Internal validation result for PICC line segmentation: Comparison of Dice score results between the existing methods and those applied with the proposed MFCN}
\label{Table3:inter_s}
\end{table}

\subsubsection{Visualization result}

\begin{figure}[htb!]
    \centering
    \includegraphics[width=0.8\textwidth]{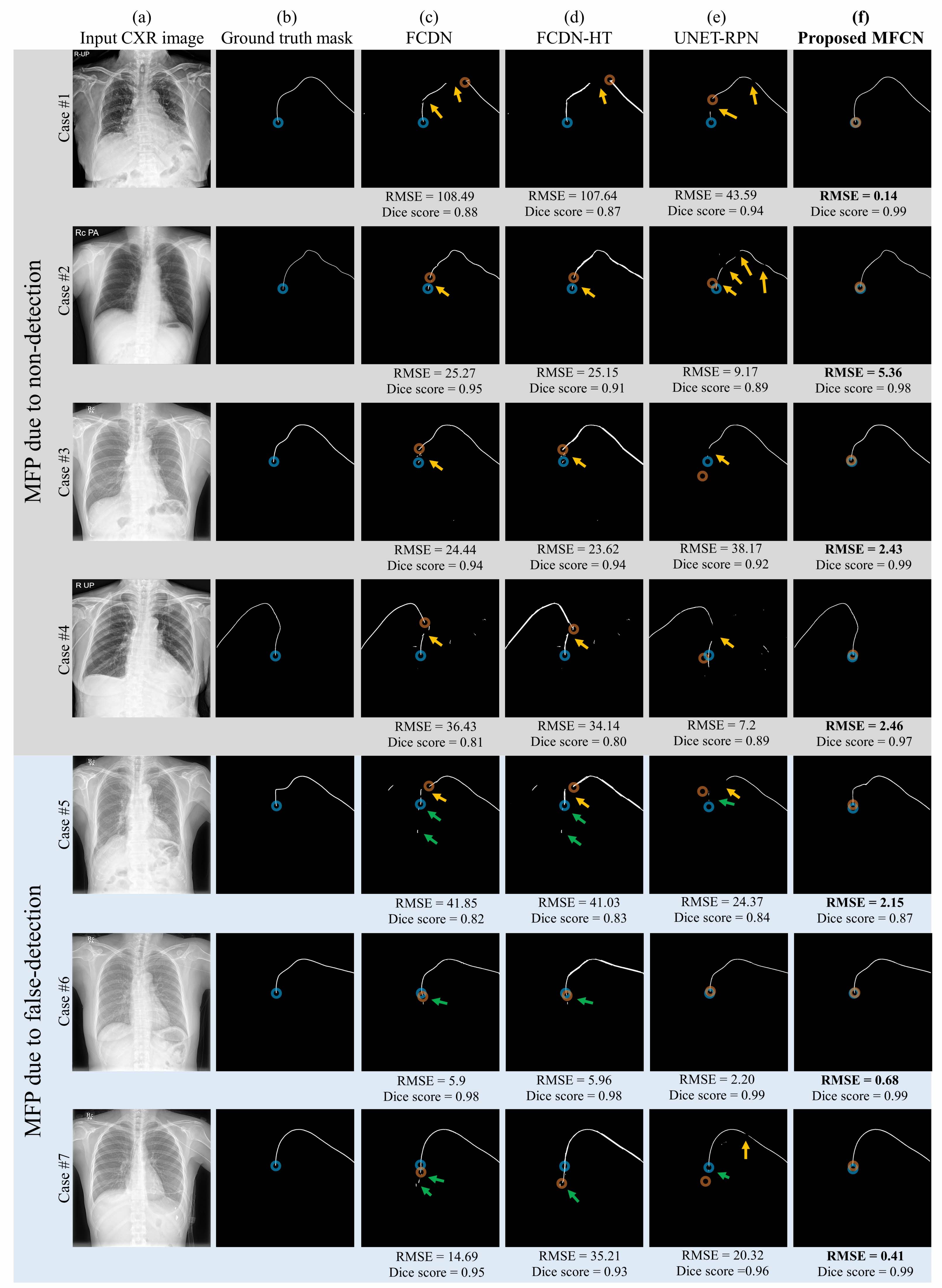}
    \caption{Internal validation examples for PICC tip detection and line segmentation through each model: (a) input CXR image, (b) ground truth PICC mask, (c)-(e) results from conventional models, (f) result from our model; The ground truth and predicted tip were marked with blue and red circle respectively. The orange arrow denotes false-negatives (i.e., when a break occurs in the line of the catheter) and the green arrow indicates false-positives (i.e., in the case of noise). The RMSE of predicted tip and Dice score of predicted PICC line were described below each image.}
    \label{fig12:inter_visual}
\end{figure}

Figure \ref{fig12:inter_visual} shows some prediction samples of the PICC line and tip through the existing model and the proposed model. It shows that accurate tip position detection was difficult in the existing model due to MFP but MFCN effectively solved it. In other words, although false-negative and false-positive of the model, which are the causes of MFP, occurred in the conventional model in Figures \ref{fig12:inter_visual}(c)-(e) (i.e., orange and green arrows respectively), the proposed method effectively solves this problem and accurately detects the tip position. Compared to the existing FCDN, the existing FCDN-HT decreased false-negatives in a small area (e.g., the first case in Figures \ref{fig12:inter_visual}(c) and (d)), but the disconnection was not resolved when false-negative appeared in a large area (e.g., the first and fourth cases in Figures \ref{fig12:inter_visual}(c) and (d)). In addition, it is observed in some cases (e.g., the seventh case in Figures \ref{fig12:inter_visual}(c) and (d)) that FCDN-HT increases false-positive so RMSE compared to FCDN. On the other hand, when our MFCN was applied, false-negative (e.g., the first and fourth cases in Figure \ref{fig12:inter_visual}(f)) and false-positive (e.g., the seventh case in Figure \ref{fig12:inter_visual}(f)) of a wide area were effectively improved, and finally RMSE for PICC tip location detection was significantly reduced as shown in Table \ref{Table2:inter_r}. 

\subsection{External validation result}
\subsubsection{PICC tip location detection} \label{Sec:external_tip_detection}

In Sections \ref{Sec:internal valid}, we validated our technique with internal data from our institution. For more objective performance verification, this section compared and evaluated the performance of the proposed MFCN and the existing technologies by using the public data (i.e., RANZCR) as introduced in Section \ref{Sec:external_data} for external verification. 

As the image in RANZCR dataset is originally given in the form of a JPG image rather than a DICOM, it is difficult to determine the pixel spacing value, so we measured RSME in pixels of JPG image not mm. Accordingly, it was impossible to calculate the frequency of samples with RMSE within 1cm so that we omitted to show this result in this experiment. 

The quantitative performance comparison results were presented in Table \ref{Table4:external_r}. Similarly with internal validation results, we also observed that MFCN consistently improved each baseline model even in this external validation, by reducing the mean RMSE by more than half (i.e., the first column in Table \ref{Table4:external_r}) and by increasing no MFP rate by more than 10\% and achieving more than 90\% (i.e., the second column in Table \ref{Table4:external_r}). These results assisted in demonstrating the objective superiority of the proposed MFCN.

\begin{table}[hbt!]
\footnotesize
\centering
\resizebox{0.7\textwidth}{!}{%
\begin{tabular}{ccccc}
\hline
 & \multicolumn{2}{c}{RMSE (mean±sd, pixels)} & \multicolumn{2}{c}{No MFP (\%)} \\ \cline{2-5} 
\multirow{-2}{*}{\begin{tabular}[c]{@{}c@{}}Conventional\\ Model name\end{tabular}} & Baseline & \textbf{MFCN} & Baseline & \textbf{MFCN} \\ \hline 
FCDN \cite{jegou2017one} & 95.90 ± 173.21 & \textbf{43.06 ± 88.70} & 71 & \textbf{96} \\
UNET \cite{ronneberger2015u} & 263.70 ± 299.97 & \textbf{71.90 ± 142.03} & 43 & \textbf{91} \\
AUNET \cite{oktay2018attention} & 187.98 ± 243.62 & \textbf{56.64 ± 126.43} & 46 & \textbf{94} \\
FCDN-HT \cite{lee2018deep} & 89.80 ± 161.768 & \textbf{49.32 ± 99.86} & 85 & \textbf{96} \\
UNET-RPN \cite{yu2020detection} & 177.98 ± 217.105 & \textbf{63.96 ± 153.33} & 67 & \textbf{95} \\ \hline
Average & 163.07 & \textbf{56.98} & 62.4 & \textbf{94.4} \\ \hline
\end{tabular}%
}
\caption{External validation result for PICC tip location detection: Comparison of RMSE results between the existing methods and those applied with the proposed MFCN}
\label{Table4:external_r}
\end{table}

\subsubsection{Visualization result} \label{Sec:external_visual}

\begin{figure}[htb!]
    \centering
    \includegraphics[width=0.8\textwidth]{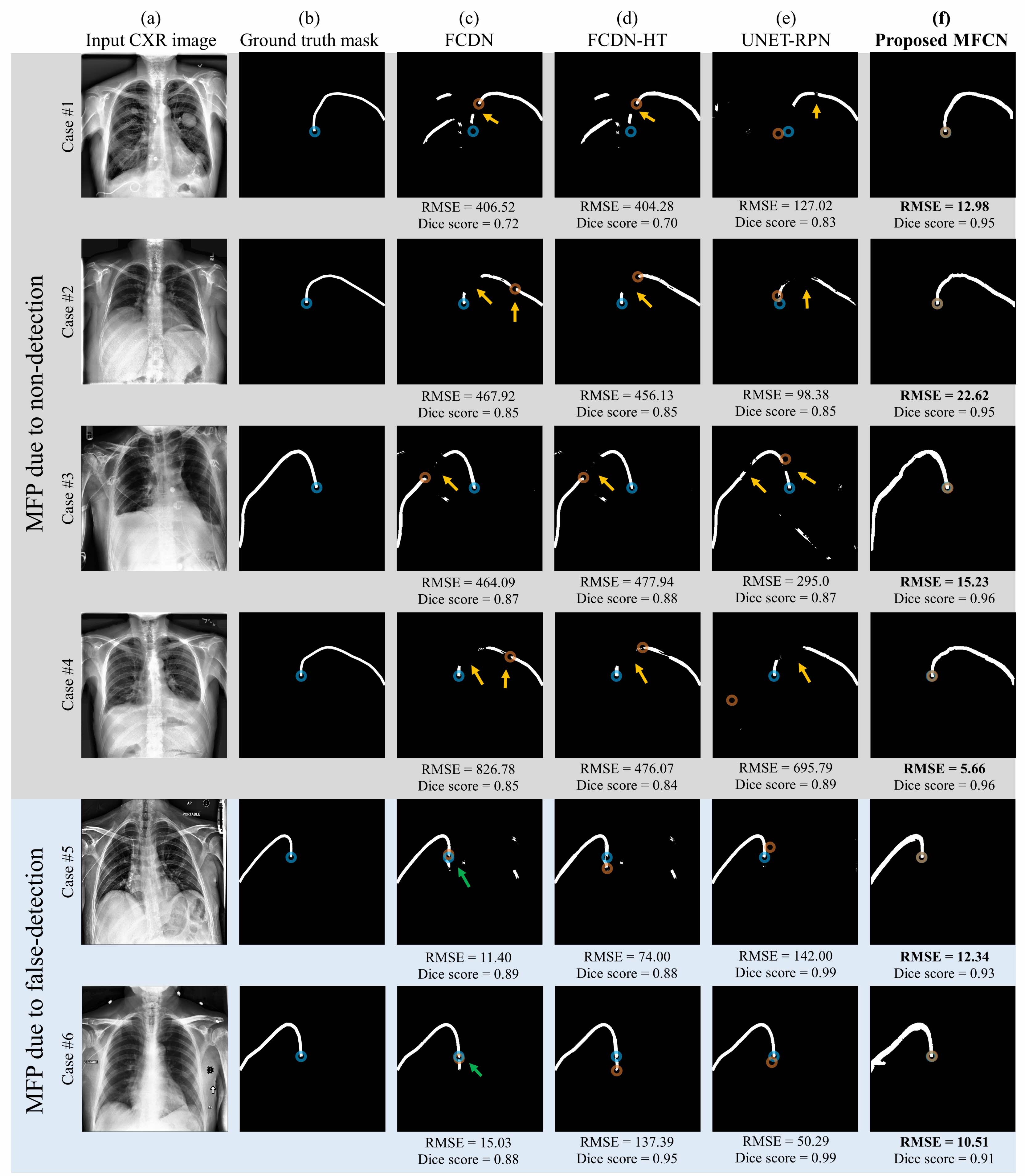}
    \caption{External validation examples for PICC tip detection and line segmentation through each mode}
    \label{fig13:exter_visual}
\end{figure}

We also showed in Figure \ref{fig13:exter_visual} model prediction examples of the PICC line and tip through the existing model and the proposed model in this external validation. Similarly with the results of internal validation, it is generally observed in Figure \ref{fig13:exter_visual} that false-negative and false-positive, the cause of MFP occurrence, appeared the same in external validation of the existing model. In the external verification, it was observed in some cases (e.g., the fifth and sixth cases in Figure \ref{fig13:exter_visual}) that the RMSE of existing FCDN-HT increased compared to its previous version (FCDN) in the case of false-positive. On the other hand, when our MFCN is applied in FCDN, the false-positive is removed so the tip position is not extended incorrectly (e.g., the fifth and sixth cases in Figure \ref{fig13:exter_visual}(f)), and the false-negative of a wide area is also effectively improved (e.g., the second and fourth cases in Figure \ref{fig13:exter_visual}(f)). As such, it can be confirmed through those examples that the proposed MFCN effectively resolves the MFP and thereby reduces the RMSE of PICC tip detection in large margins as shown in Table \ref{Table4:external_r}.

\section{Discussion}
\label{Sec:Discussion}

\subsection{Causes of MFP and how to solve them through our work}
Typically, the causes of MFP can be classified into the following four categories:

For the first reason, as the CXR image is taken by projecting the 3D structure in 2D, a part of the catheter may be obscured by the anatomical structure \cite{yi2020automatic}. If the catheter overlaps with the anatomical structure, the catheter edge can be indistinguishable as the bone and the catheter’s pixel values are similar, as shown in Figure \ref{fig1:MFP} \cite{yu2020detection}. Then, a lot of prediction errors can occur from such overlapping parts. Specifically, it is confused with a bone having a linear structure partially similar to PICC, such as a spine or rib, and is incorrectly predicted as a catheter \cite{keller2007semi}. 

As PICC is a very thin tube structure with an average pixel-level thickness ranged from 3 to 6, it is very sparse occupying on average only about 0.14\% of the total CXR image. Due to such sparsity or pixel-level class imbalance for segmentation, it is easy to treat the entire portion corresponding to the sparse area (e.g., PICC) as noise or to classify it as background pixels \cite{greenland2016sparse}. There exists an attempt to address this issue by adopting the class-balanced cross-entropy (BCE) loss function \cite{xie2015holistically}. However, even if the class-balanced loss is considered, the PICC estimate can be easily divided into multi-fragments due to false negatives for some pixels, making it difficult to determine the exact location of the tip. 

For the third reason, using the full CXR image as input makes it difficult to extract the PICC line, decreasing the accuracy for extracting the PICC. It is worth noting that segmentation task requires extensive down sampling or pooling within the internal path of CNN. This characteristic makes the latent features of PICC easily removed, especially when the network input takes the whole CXR image with original resolution. It is because the PICC line is relatively sparse compared to the entire CXR image.

For the last reason, X-ray images are expensive to acquire and manual labeling is time-consuming, making it difficult to acquire large and diverse databases \cite{bullock2019xnet, sharma2010automated}. In order to extract a more accurate PICC line, the label data obtained by accurately extracting PICC lines from training data should be required to train the model. However, the tip detection and masking process is difficult and time consuming because the PICC tip is in many cases obscured by the shadow of the spine on the CXR images. Though reducing the training data can be regarded as an alternative, in this case the trained model is highly overfitted, decreasing the accuracy of PICC line extraction in the actual diagnostic test.

\subsection{Related works for PICC line and tip detection}

PICC tip prediction studies for early detection of movement have been conducted using various techniques such as CXRs \cite{wechsler1984misplaced, harako2004optimizing, bailey2000immediate}. electrocardiogram (ECG) \cite{gao2018safety, oliver2016ecg}, and ultrasound \cite{moureau2003using, nicholson2010development}, among which CXRs were used as a gold-standard \cite{li2018randomized, black2000central}. 

In particular, interest in computer-aided detection (CAD) for the PICC tip location in CXR images is increasing significantly. It is because computer-aided detection (CAD) can be used to help radiologists interpret medical images and reduce mistakes, and its performance has dramatically improved due to recent advances in artificial intelligence.

Several traditional CADs for segmenting the catheter have been proposed. As a representative technique, Keller \textit{et al} proposed a semi-automated system for detecting catheters on CXRs \cite{keller2007semi}. By quantitatively analyzing the object intensity between the catheter line and the tip through this method through image processing and pattern recognition techniques, it was possible to secure the position of the tip more accurately than the existing methods. However, this method has a limitation in that it is difficult to perform pattern analysis on the intensity of the data because the profile of the object is different depending on the X-ray intensity and the imaging environment. It was also reported that this algorithm has problems such as stopping when the PICC overlaps with anatomical structures, mentioned as the first cause of MFP.

Recent advances in DL technology applied to medical imaging have shown great potential for increasing diagnostic accuracy and image interpretation speed, and these DL-based approaches also have been introduced in tip and line detection for PICC misposition diagnosis. 

Henderson \textit{et al} proposed a method to accurately detect the presence and type of catheters, and as a result, they succeeded in classifying the presence or absence of four catheters of interest (NGT, ETT, UAC, UVC) on radiographs of the chest and abdomen of newborns with 95\% precision \cite{henderson2021automatic}. However, though their technique efficiently detects the presence of catheter and the type of catheter, it does not provide the position of the PICC tip, so there is a limitation in that the expert clinicians can directly diagnose the misposition from this CAD result. 

In order to overcome this limitation, there have been attempts to directly extract PICC lines based on DL. Subramanian \textit{et al} used a large-scale CVC data, classified four common CVC types by using random forest, and roughly identified their potential fragments through a UNET-based model \cite{subramanian2019automated}. However, their results were an initial study to obtain the approximate shape of the PICC line, and they did not try to directly improve the MFP problem as in our study to make tip detection precise. 

\subsection{Related works for MFP improvement}

Recent studies have attempted to improve MFP to accurately detect PICC tip detection based on predicted segmentation results. 

Lee \textit{et al} proposed a cascading segmentation AI system containing two fully convolutional neural networks to improve MFP and detect tips more accurately \cite{lee2018deep}. Their cascading system tried to improve the catheter tip detection accuracy by supplementing the false-negative through the Hough line transform algorithm. However, as they reported, their system still did not perform perfectly in cases that could be confused with similar appearing artifacts or structures (e.g., edges of bone structures) with PICC, suggesting the need for additional research resolving the corresponding false-positives \cite{lee2018deep}. Specifically, as shown in experiments in Appendix, we found that it was difficult for their algorithm to effectively solve the MFP problem of various CXR images because the parameters of the Hough line transformation for correcting the PICC line were different for each CXR input image. Our method mainly differs from their algorithm in that we additionally improve the false-positives by proposing a DL-based MFP improvement technique (i.e., the third stage network for line reconnection). Compared to Hough line transform-based one \cite{lee2018deep}, our DL-based approach experimentally further improves FPs and thus the tip detection performance without being sensitive to parameters like the Hough transform, verifying the effectiveness of our technique.

Ambrosini \textit{et al} proposed a technique for correcting the MFP by connecting the negative areas between two positive points in the segmentation prediction result when the negative area is sufficiently small \cite{ambrosini2017fully}. However, this technique does not compensate for a wide undetected area and cannot correct for false detection, suggesting the need for additional research \cite{ambrosini2017fully}. Based on the higher radiation dose fluoroscopy image as used in \cite{ambrosini2017fully}, there was almost no false-positive. However, as in the case of the more normal or low-dose CXR as assumed in our study, the more false-positive of the network may occur (e.g., Figure \ref{fig1:MFP}), so improvement should be considered when solving the MFP. The proposed study is mainly different from the previous research \cite{ambrosini2017fully} in that our method is able to further resolve even a large undetected area thanks to the technical characteristics, and it also improves the false-positive at the same time, by letting the network recognize the position of tip, which is the end point of PICC, together when learning for MFP calibration.

Yu \textit{et al} also detected the tip position more precisely by performing object detection on the tip position separately in addition to PICC line segmentation \cite{yu2020detection}. However, their tip position detection method through object detection was designed to be affected by the PICC line segmentation result, so the inaccuracy of the PICC line segmentation result affects the performance of their object detection-based tip position detection. In this aspect, resolving the MFP problem of PICC line segmentation can improve tip detection performance, but their results did not take this into account. Focusing on these issues, we reconstructed their results, applied the proposed technology, and confirmed the performance improvement, verifying the effectiveness of our work in comparison with their method. 

\subsection{Limitation of our study}

This study has the following three limitations. First, the CXR images collected for internal validation were limited to only low-dose images so may generate various kinds of noise, though we supplemented it through external validation. Second, unlike the existing studies using portable X-rays \cite{yu2020detection, lee2018deep}, our study was conducted using Chest PA. Compared to portable X-rays, chest PA may have relatively more fixed positions of the shoulder and heart and clear image. For this reason, the learning model through chest PA may show different results from the existing studies conducted with portable X-rays. However, we expected that our method will be effective even in portable X-ray as well as it was not limited to PA but general types in external validation. Lastly, for simplicity, our study was conducted with images containing only PICC. In other words, the proposed study was focused on the possibility that there could be a performance improvement compared to the existing technology, so the data were limited to data with only PICC. However, the tip detection accuracy of all models may be degraded when various types of catheters are taken simultaneously in the CXR image. Therefore, a study to evaluate the performance of the proposed and other techniques may be of interest as a follow-up study in the case of CXR images including a catheter other than PICC.

\section{Conclusions}
\label{Sec:Conclusions}

In this study, we propose a multi-stage network to improve the PICC tip detection performance of the existing DL-based models by solving the MFP occurring in their segmentation results. To achieve it, the proposed scheme is designed to add two networks to the conventional model for PICC line segmentation. The first added model affects as accurately extracting PICC line with little data by allowing the network to effectively focus on the sparsely expressed PICC area in the CXR image. The second added model was designed to recognize and restore noise (false-positive) confusing with the actual PICC line and the unexpected PICC line breakage (false-negative) due to partially obscured by anatomical structures or weakly expressed edges. With extensive experiments through internal validation and external validation, we verified our work by showing that it consistently improves tip position detection performance of existing models by more than 63\%. We also strengthened the clinical significance by showing the validity of this study in the migrated data. These all aspects suggest that our fully-automatic PICC tip localization approach can play a high auxiliary role in detecting abnormal migration of the PICC tip by doctors, further contributing to the prevention of complications.

\appendix

\section{Appendix: Additional experiments}
\label{Sec:Appendix}

\subsection{Performance comparison experiment for combination of individual stage modules in proposed MFCN} 
\label{Sec:stage compare}

In order to prove the validity of the proposed module for each stage in our MFCN, the RMSE performance with each combination of individual stage modules was compared in Table \ref{Table5:ablation}. This result shows that the RMSE performance gradually improved whenever the proposed model was added from stage 1 to 3. In the case of the second stage module, it did not show a significant performance improvement when combined with the first stage module, but showed a meaningful performance improvement when the second stage module was added to the combined model of the first and third modules (i.e., RMSE was reduced by more than 30\% from 14.21 to 9.37 mm). Therefore, we demonstrated from these results that considering both the second stage and third stage models simultaneously was effective for PICC tip localization.

\begin{table}[hbt!]
\footnotesize
\centering
\resizebox{0.7\textwidth}{!}{%
\begin{tabular}{cccc}
\hline
\multicolumn{3}{c}{Ablation configuration} & \multirow{2}{*}{RMSE (mean±sd, mm)} \\ \cline{1-3}
First stage model & Second stage model & Third stage model &  \\ \hline
\checkmark & w/o & w/o & 23.37 ± 32.12 \\
\checkmark & \checkmark & w/o & 21.56 ± 33.24 \\
\checkmark & w/o & \checkmark & 14.21 ± 23.21 \\
\checkmark & \checkmark & \checkmark & 9.37 ± 18.13 \\ \hline
\end{tabular}%
}
\caption{Ablation results of proposed MFCN: RMSE results for tip location detection (mean±sd, mm) were given by internal validation data}
\label{Table5:ablation}
\end{table}

\subsection{Performance evaluation of the proposed MFCN model in the misposition case} \label{Sec:misposition}

In the previous section \ref{Sec:internal valid}, we have demonstrated the superiority of the PICC tip position detection of the proposed technique in general regardless of misposition (i.e., for evaluation data in which both misposition cases and non-misposition cases are mixed). In this section, we additionally showed the performance comparison results between the proposed MFCN and the existing technologies only for misposition case. In other words, only the case of misposition was left for our test data in the same environment as shown in Section \ref{Sec:internal valid}. As shown in Table \ref{Table1:dataset}, there were 13 misposition CXR samples in the test data so we evaluated only these 13 data.

We presented the performance comparison results for this misposition case between the proposed MFCN and the existing technologies for PICC tip location detection in Table \ref{Table6:internal_mispo} as the same form of Table \ref{Table2:inter_r}. Similarly with results of Table \ref{Table2:inter_r}, Table \ref{Table6:internal_mispo} shows that MFCN consistently improved each baseline model even in this misposition case, reducing the mean RMSE by more than half, increasing no MFP rate by more than 40\%, and increasing the rate of accurate PICC tip location detection (i.e., RMSE less than 1cm) by more than 20\% for all comparisons with baseline models. 

\begin{table}[hbt!]
\footnotesize
\centering
\resizebox{0.7\textwidth}{!}{%
\begin{tabular}{ccccccc}
\hline
 & \multicolumn{2}{c}{RMSE (mean ± sd, mm)} & \multicolumn{2}{c}{No MFP (\%)} & \multicolumn{2}{c}{$|RMSE|$ \textless 1cm (\%)} \\ \cline{2-7} 
\multirow{-2}{*}{\begin{tabular}[c]{@{}c@{}}Conventional\\ Model name\end{tabular}} & Baseline & \textbf{MFCN} & Baseline & \textbf{MFCN} & Baseline & \textbf{MFCN} \\ \hline
FCDN \cite{jegou2017one} & 39.73 ± 42.37 & \textbf{11.7 ± 15.66} & 0 & \textbf{85} & 31 & \textbf{69} \\
UNET \cite{ronneberger2015u} & 32.01 ± 36 & \textbf{10.54 ± 12.12} & 46 & \textbf{85} & 46 & \textbf{69} \\
AUNET \cite{oktay2018attention} & 44.34 ± 44.79 & \textbf{11.46 ± 14.63} & 31 & \textbf{85} & 31 & \textbf{69} \\
FCDN-HT \cite{lee2018deep} & 31.23 ± 42.24 & \textbf{11.64 ± 15.58} & 15 & \textbf{85} & 46 & \textbf{69} \\
UNET-RPN \cite{yu2020detection} & 23.52 ± 21.3 & \textbf{10.86 ± 12.83} & 0 & \textbf{85} & 46 & \textbf{69} \\ \hline
Average & 34.17 & \textbf{11.24} & 18.4 & \textbf{85} & 40 & \textbf{69} \\ \hline
\end{tabular}%
}
\caption{Internal validation result for PICC tip location detection (only with misposition data): Comparison of RMSE results between the existing methods and those applied with the proposed MFCN}
\label{Table6:internal_mispo}
\end{table}

\begin{figure}[htb!]
    \centering
    \includegraphics[width=0.8\textwidth]{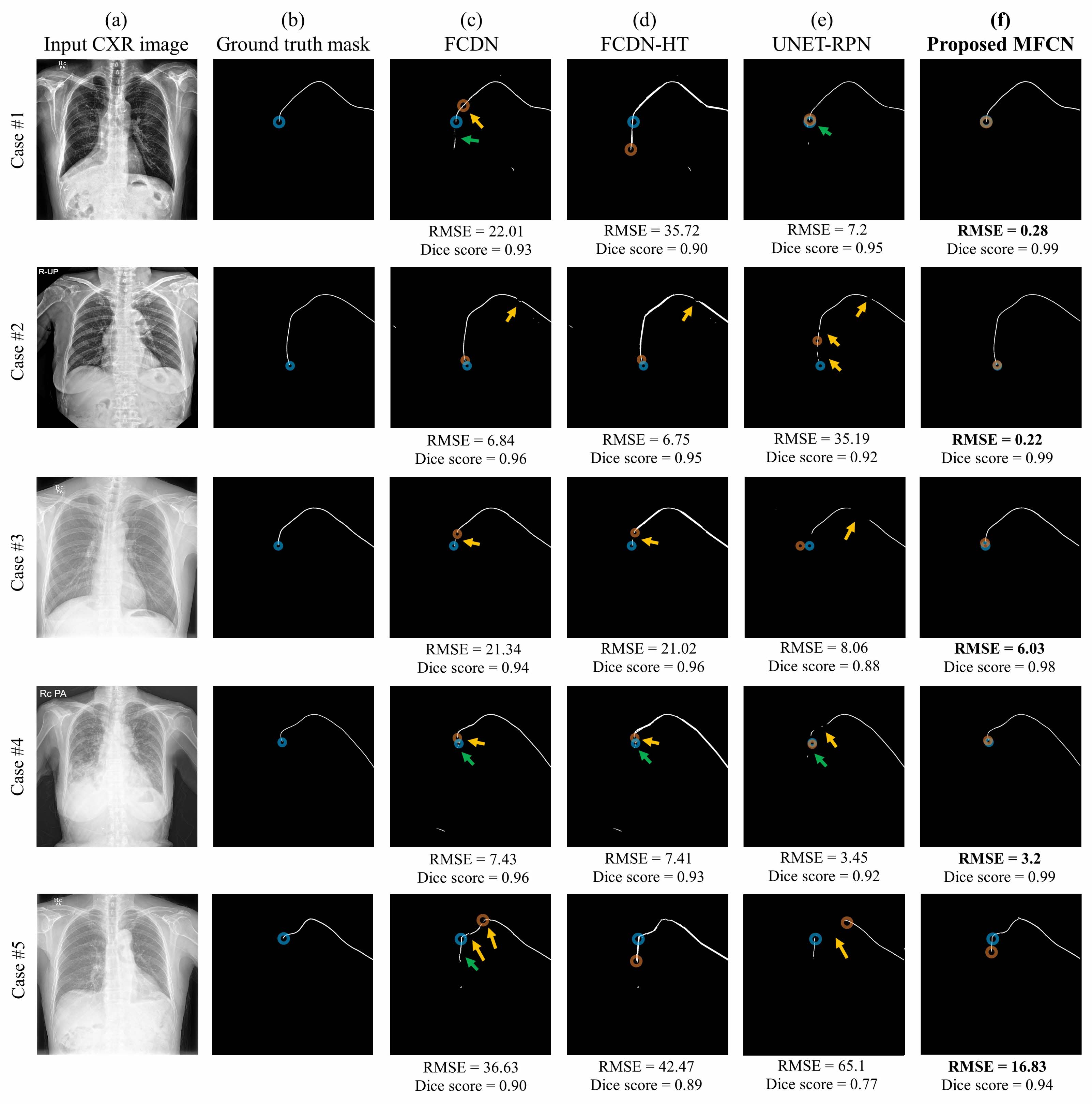}
    \caption{Internal validation examples of misposition data for PICC tip detection and line segmentation through each model: (a) input CXR image, (b) ground truth PICC mask, (c)-(e) results from conventional models, (f) result from our model; The ground truth and predicted tip were marked with blue and red circle respectively. The orange arrow denotes false-negatives (i.e., when a break occurs in the line of the catheter) and the green arrow indicates false-positives (i.e., in the case of noise). The RMSE of predicted tip and Dice score of predicted PICC line were described below each image.}
    \label{fig14:inter_mis}
\end{figure}

Figure \ref{fig14:inter_mis} shows the prediction examples of the PICC line and tip through the existing model and the proposed model. It shows that MFP generated from mispositioned data degraded accurate tip detection performance. Specifically, as the mispositioned PICC has a different tip position from the normal position, it can cause confusion with the spine, resulting in false-positive in the prediction of the existing model (e.g., the first and fifth cases in Figure \ref{fig14:inter_mis}). It was also confirmed that there was difficulty in distinguishing false negatives due to misposition (e.g., the third case in Figure \ref{fig14:inter_mis}). It is also observed that the existing UNET-RPN incorrectly predicts the tip position as normal position (e.g., the second case in Figure \ref{fig14:inter_mis}). On the other hand, when the proposed MFCN was applied, MFP is effectively solved so that RMSE for PICC tip location detection was reduced consistently for all existing models and samples (e.g., Figure \ref{fig14:inter_mis}(f)), thanks to our post-processing of MFCN.

\subsection{Backbone network for proposed MFCN} \label{Sec:backbone}

It is useful to review that any existing network for segmentation can be used as backbone network for the second stage and third stage models in proposed MFCN. In Table \ref{Table7:backbone}, nine RMSE results of MFCN for tip position detection were compared by applying three existing backbone (i.e., FCDN, UNET, and AUNET) models to each of the second stage and third stage models. All nine cases of the proposed MFCN (the first/second/third stages) showed the performance improvement compared to the conventional model (the first stage, RMSE = 23.37 ± 32.12 mm), confirming that the proposed MFCN has improved performance regardless of the specific backbone model. In this study, we selected the FCDN as our second and third stage model as it performs the highest performance as shown in Table \ref{Table7:backbone}.

\begin{table}[hbt!]
\footnotesize
\centering
\resizebox{0.7\textwidth}{!}{%
\begin{tabular}{cccc}
\hline
\multirow{2}{*}{Second stage model} & \multicolumn{3}{c}{Third stage model} \\ \cline{2-4} 
 & FCDN \cite{jegou2017one} & UNET \cite{ronneberger2015u} & AUNET \cite{oktay2018attention} \\ \hline
FCDN \cite{jegou2017one} & \textbf{9.37 ± 18.13} & 14.24 ± 24.35 & 16.26 ± 27.59 \\
UNET \cite{ronneberger2015u} & 14.20 ± 21.37 & 16.59 ± 22.53 & 18.55 ± 25.15 \\
AUNET \cite{oktay2018attention} & 15.99 ± 21.35 & 14.24 ± 24.35 & 16.26 ± 27.59 \\ \hline
\end{tabular}%
}
\caption{RMSE (mean±sd, mm) of MFCN for tip location detection with different backbone model: The first stage (conventional) model was set to FCDN (with RMSE equal to 23.37 ± 32.12 mm). The second and third stage models were set as follows. The RMSE results were measured in the internal validation dataset.}
\label{Table7:backbone}
\end{table}

\subsection{Performance with varying Hough transform parameters} \label{Sec:Hough}
In the case of existing segmentation network FCDN-HT, performance changes depending on the parameter settings of the Hough transform so that we additionally compared it in Table \ref{Table8:hough}. The Hough transform mainly consists of the following three parameters, the minimum number of intersecting points to detect a line (MIP), minimum length of line (MLL), and maximum allowed gap between line segments for a single line (MLG). We changed these HT parameters in 5 units between the value regions between 5 and 80 and presented the optimal parameter candidate group in Table \ref{Table8:hough}. Specifically, we set MIP/MLL/MLG to 5/80/70 and 5/20/5 as they provided the best RMSE in test and validation dataset respectively. We also set MIP/MLL/MLG to 70/30/5 as it showed the best Dice score in test dataset. Finally, we set MIP/MLL/MLG to 50/30/50, following the original suggestion of FCDN-HT \cite{lee2018deep}. As a result of the RMSE analysis in Table \ref{Table8:hough}, compared to FCDN, FCDN-HT reduced the RMSE value by an average of 6.55 mm (2.35 mm to 11.09 mm) in all HT parameters, but still our MFCN showed the best tip localization performance (i.e., RMSE = 9.37mm) than in all cases of FCDN-HT (i.e., RMSE larger than 12.28 mm).  

\begin{figure}[htb!]
    \centering
    \includegraphics[width=0.8\textwidth]{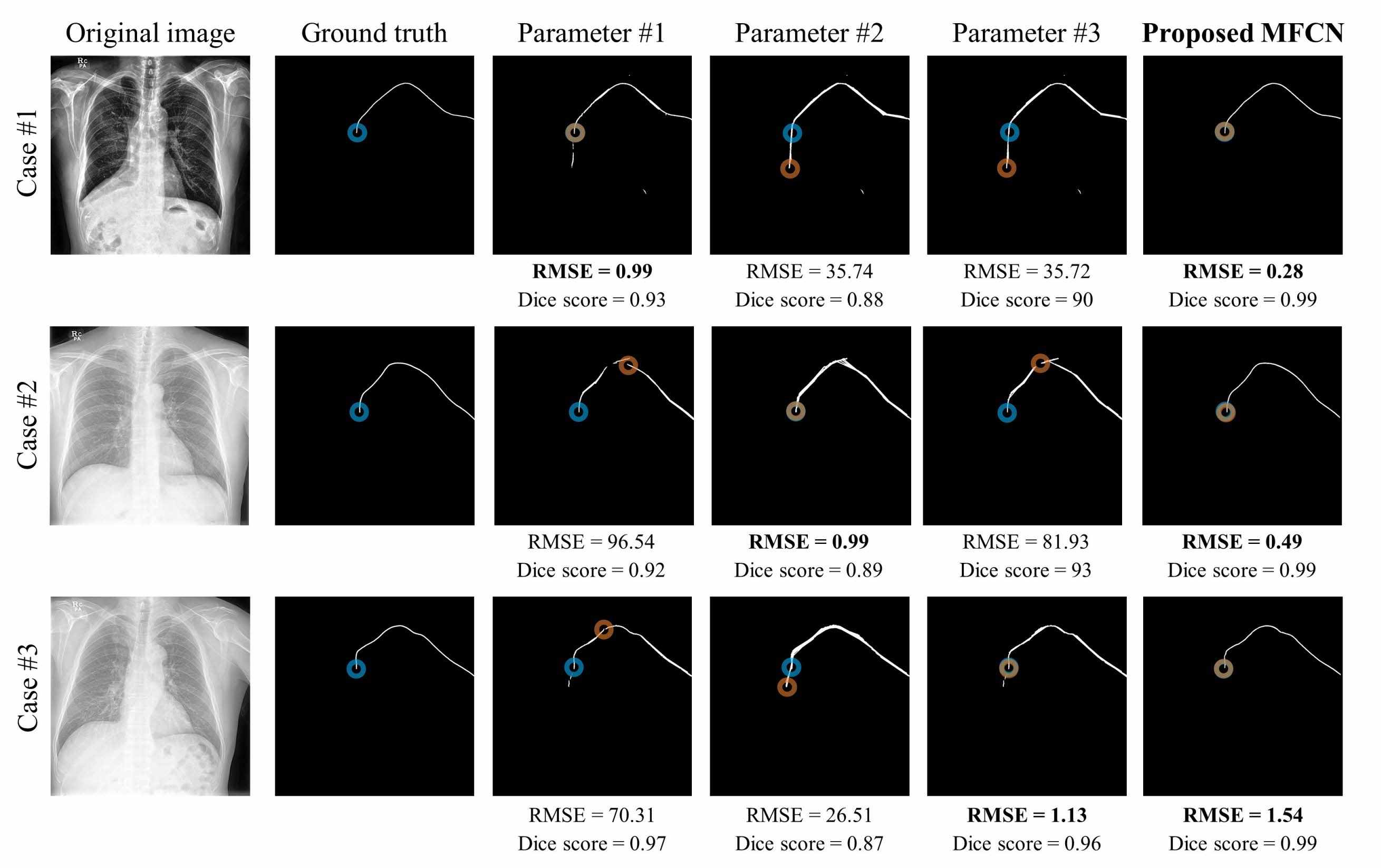}
    \caption{Examples for PICC line extraction and tip location detection through proposed MFCN and existing FCDN-HT with different Hough transform parameters: Parameters \#1, \#2, and \#3 denote the cases of MIP/MLL/MLG equal to (70/30/5), (5/80/70), and (50/30/50), respectively.}
    \label{fig15:hough}
\end{figure}

In addition to the quantitative performance comparison in Table \ref{Table8:hough}, the qualitative performance results of PICC tip prediction images were presented in Figure \ref{fig15:hough}. These show that the optimal HT parameter of FCDN-HT is different for each CXR image due to the difference in the distance and shape of FN (i.e., in case of generating a break in the line estimate) and FP (i.e., in the case of generating noise). However, our MFCN shows better or similar tip position detection performance without parameter influence.

\begin{table}[hbt!]
\footnotesize
\centering
\resizebox{0.7\textwidth}{!}{%
\begin{tabular}{ccc}
\hline
 & RMSE (mean±sd, mm) & Dice score (mean±sd, mm) \\ \hline
FCDN & 23.37 ± 32.12 & 0.91 ± 0.16 \\ 
FCDN-HT (5 / 80 / 70) & 12.28 ± 22.19 & 0.86 ± 0.13 \\
FCDN-HT (5 / 20 / 5) & 19.63 ± 29.62 & 0.90 ± 0.14 \\
FCDN-HT (70 / 30 / 5) & 21.02 ± 31.1 & 0.91 ± 0.15 \\
FCDN-HT (50 / 30 / 50) \cite{lee2018deep} & 17.16 ± 28.25 & 0.90 ± 0.14 \\ 
\textbf{Proposed MFCN} & \textbf{9.37 ± 18.13} & \textbf{0.92 ± 0.17} \\ \hline
\end{tabular}%
}
\caption{Comparison of internal validation RMSE results by varying hough transform parameters (MIP/MLL/MLG) in FCDN-HT}
\label{Table8:hough}
\end{table}

\subsection{Validation with different training data size} \label{Sec:training data}

\begin{figure}[htb!]
    \centering
    \includegraphics[width=0.7\textwidth]{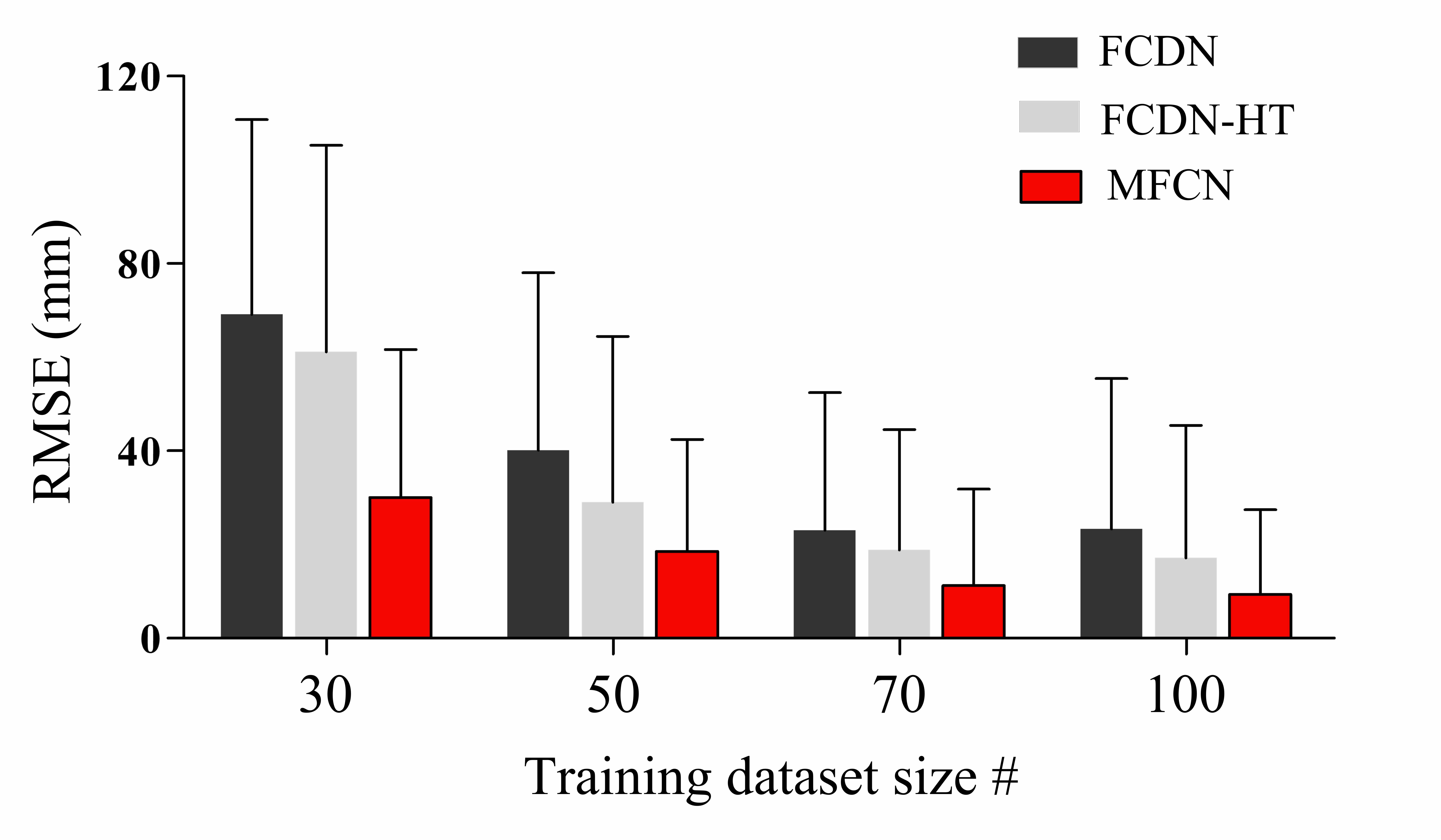}
    \caption{Internal validation RMSE result of PICC tip localization with varying number of training dataset: proposed MFCN is applied to the baseline FCDN for simplicity}
    \label{fig16:train_datasize}
\end{figure}

In this section, we compared the proposed MFCN performance with other models on a smaller training dataset. Specifically, we varied the number of training data as 100, 70, 50, and 30 and provided these results in Figure \ref{fig16:train_datasize}. When the proposed MFCN is applied to the baseline FCDN model, it consistently improves the tip location detection performance with RMSE reduced by more than 50\% than that of the target baseline MFCN. The accurately labeled X-ray data is required to train the PICC detection model, but it is difficult to acquire a large and diverse database due to the high labeling and data correction costs \cite{bullock2019xnet, sharma2010automated}. In this respect, the experimental results demonstrated that the proposed technique showed high PICC tip detection performance from less training data than before, proving its high practicality.

\subsection{Multi-fold evaluation} \label{Sec:monte}

In all experiments so far, for simplicity, we performed holdout/one-fold validation with a sufficient inference data set size. In this section, we also proceed with additional validation by using 5-fold Monte Carlo validation to observe whether there exists a meaningful difference with the proposed holdout validation. Specifically, 150 test data were randomly selected from 280 data given for each of internal and external validation, and 100 and 30 were selected as training and validation data from the remaining data, respectively. Then the model evaluation was repeated 5 times independently and the mean and variance values were included in Table \ref{Table9:carlo}. As a result, MFCN reduced the RMSE of the existing model (i.e., MFCN) to less than half in both internal and external validations, as similarly with the hold-out validation results. This implies that the superiority of the proposed technology can be demonstrated sufficiently with holdout validation. 

\begin{table}[hbt!]
\footnotesize
\centering
\resizebox{0.7\textwidth}{!}{%
\begin{tabular}{ccccc}
\hline
 & \multicolumn{2}{c}{\begin{tabular}[c]{@{}c@{}}Internal validation RMSE\\(mean±sd, mm)\end{tabular}} & \multicolumn{2}{c}{\begin{tabular}[c]{@{}c@{}}External validation RMSE\\(mean±sd, pixels)\end{tabular}} \\ \cline{2-5} 
\multirow{-3}{*}{\begin{tabular}[c]{@{}c@{}}Conventional\\ Model name\end{tabular}} & Baseline & \textbf{MFCN} & Baseline & \textbf{MFCN} \\ \hline
FCDN & 18.99 ± 26.28 & \textbf{9
05± 19.12} & 95.64± 145.56 & \textbf{43.16 ± 95.16} \\ \hline
\end{tabular}%
}
\caption{Internal validation RMSE result of PICC tip localization with varying number of training dataset: proposed MFCN is applied to the baseline FCDN for simplicity}
\label{Table9:carlo}
\end{table}



\section{Acknowledgements}

This work was supported by the MSIT (Ministry of Science and ICT), Korea, under the ITRC (Information Technology Research Center) support program (IITP-2021-2018-0-01798) supervised by the IITP (Institute for Information \& Communications Technology Planning \& Evaluation). This work was also supported by the Future Medicine 20*30 Project of the Samsung Medical Center (SMX1210791).

\clearpage 
\bibliographystyle{unsrtnat}
\bibliography{reference.bib} 
\clearpage

\end{document}